\documentclass{JHEP3}
\usepackage{epsfig, amssymb, bm, graphicx, amsmath}

\usepackage{color}
\def\be{\begin{equation}}
\def\ee{\end{equation}}    
\def\ba{\begin{eqnarray}}
\def\ea{\end{eqnarray}}

\def\lsim{\mbox{\raisebox{-.6ex}{~$\stackrel{<}{\sim}$~}}}
\def\gsim{\mbox{\raisebox{-.6ex}{~$\stackrel{>}{\sim}$~}}}

\def\2pi{\left(2\pi\right)}

\def\nn{\nonumber \\}

\def\grad{\vec{\nabla}}

\newcommand{\Expect}[1]{\left\langle #1 \right\rangle}

\title{Feeding your Inflaton: Non-Gaussian Signatures of Interaction Structure}

\author{Neil Barnaby\\
School of Physics and Astronomy, University of Minnesota\\
116 Church Street S.E., Minneapolis, MN USA\\
E-mail: \email{barnaby@physics.umn.edu}}

\author{Sarah Shandera\\
Dept.~of Physics, Pennsylvania State University\\
104 Davey Lab, University Park, PA USA\\
E-mail: \email{shandera@gravity.psu.edu}
}

\abstract{
Primordial non-Gaussianity is generated by interactions of the inflaton field, either self-interactions or couplings to other sectors. These two physically different mechanisms can lead to nearly indistinguishable bispectra of the equilateral type, but generate distinct patterns in the relative scaling of higher order moments. We illustrate these classes in a simple effective field theory framework where the flatness of the inflaton potential is protected by a softly broken shift symmetry. Since the distinctive difference between the two classes of interactions is the scaling of the moments, we investigate the implications for observables that depend on the series of moments. We obtain analytic expressions for the Minkowski functionals and the halo mass function for an arbitrary structure of moments, and use these to demonstrate how different classes of interactions might be distinguished observationally.  Our analysis casts light on a number of theoretical issues, in particular we clarify the difference between the physics that keeps the distribution of fluctuations nearly Gaussian, and the physics that keeps the calculation under control.
}

\preprint{IGC-11/9-2, FTPI-MINN-11/25, UMN-TH-3015/11}

\begin{document}
\maketitle

\section{Introduction}

The next few years will bring an extraordinary amount of new information about the primordial cosmological fluctuations. The Planck satellite \cite{planck} is measuring the temperature anisotropies of the Cosmic Microwave Background (CMB) with substantial improvements in precision, reaching down to smaller length scales. Current and upcoming large volume surveys are cataloging cosmological structures (eg - galaxies, clusters and voids) whose number and distribution depend on the same initial conditions that source the CMB temperature anisotropies \cite{ACT, SPT, SDSS, BOSS, DES, HETDEX, LSST, WFIRST}. These primordial fluctuations provide a remarkable observational window into the history of the early universe and offer the potential to probe fundamental physics at the energy scale of inflation, orders of magnitude above what can be reached in the laboratory. To make the most of this data, it is important to understand in detail what microscopic information is encoded in the primordial curvature fluctuations, and how this may be extracted from an observational perspective.

The increasing observational precision will allow us to significantly constrain, or even measure, statistics of the primordial fluctuations beyond the power spectrum, collectively called non-Gaussianity. To date, theoretical work and observational constraints of non-Gaussianity have focused mostly on signatures of the 3-point correlation function (the bispectrum), which provides a leading indicator of non-Gaussian effects in most scenarios. Higher order moments are challenging to compute and even more challenging to constrain observationally. But, when the primordial curvature fluctuations are calculated beginning with a particle physics model for the inflaton field(s), and {\it especially when non-Gaussianity is observably large}, there are relationships between the moments of the fluctuations coming from the structure of the original theory. This is an especially striking feature of the most robust and appealing inflationary models: they rely on some symmetry that, although weakly broken, determines much of the structure of the terms appearing in the theory. When non-Gaussianity is large, this structure can be directly mapped to the pattern of primordial fluctuations in the gravitational potential and the order-by-order perturbative expansion contains evidence of the underlying symmetry. More than the presence or absence of a three-point function, it will be consistency relations between the moments themselves (and ideally with the evolution of the inflationary background) that will make a more compelling case for inflation or force us to consider alternatives. 

In this paper we discuss two types of consistency relations that are typical of particle physics inflationary scenarios. One is a feature of 
nearly every non-Gaussian model considered in the literature so far and the other is quite new. We work, for the most part, within one of the simplest and most 
popular microscopic frameworks: inflation driven by a single scalar field with an underlying shift symmetry. Although the starting point is simple, there is a rich array of possible non-Gaussian signatures. Focusing on global properties of the correlation functions rather than just the bispectrum, we uncover two distinctive classes of interactions:
\begin{itemize}

 \item When non-Gaussianity comes from \emph{self-interactions} of a single field, the correlation functions are typically {\it hierarchical}: the $n$-th dimensionless moment scales like $\mathcal{M}_n \propto \left(\mathcal{I} \mathcal{P}_\zeta^{1/2}\right)^{n-2}$ where $\mathcal{P}_{\zeta}$ is the amplitude of the fluctuations and $\mathcal{I}$ measures the strength of the interaction ($\mathcal{I}\sim f_{NL}$).

 \item When non-Gaussianity is generated by interactions with \emph{other sectors}, the dimensionless moments exhibit a very different structure. They may fall off much more slowly, or not at all. This novel scaling of the moments indicates that such scenarios are {\it more non-Gaussian} than their hierarchical counterparts, given an amplitude of the three-point function.

\end{itemize}
We demonstrate how the two classes may be easily confused at the level of the three-point function, but would be particularly distinguished by observables sensitive to the difference in the pattern of moments, for example the statistics of collapsed objects. 

We also clarify the difference between the physics that keeps the Probability Distribution Function (PDF) nearly Gaussian, and the physics that keeps the calculation under control.  In the hierarchical case, the structure of self-interactions is intimately linked to the validity of perturbation theory for the interactions (ie, keeping loop corrections to the two-point small). In contrast, the case with non-hierarchical scaling can be under control even in a regime where the dimensionless moments are not only non-hierarchical but are not even a converging series. This case is important because it helps us to understand the theory of inflation, although the simplest models are already ruled out observationally.  

Although we will calculate features of a particular class of models, our goal is to illuminate where the programme of constraining or detecting non-Gaussianity is headed more generally. For example, a detection of non-Gaussianity from a particular bispectral template would be fantastic, but would not uniquely determine the primordial physics. We would then be faced with the need to do at least one of several difficult measurements: a very precise measurement of the bispectral shape, a measurement of the four point function, or a measurement that would fold in information about higher order moments. The first two are technically challenging but straightforward, so it is the last option that we will explore in this paper. 

We will find that our analysis is also relevant for a more general issue: any constraint or detection of non-Gaussianity should be confirmed with observables that are sensitive to non-Gaussianity but that are not direct $n$-point function measurements. For example, we might cross-check measurements of Minkowski functionals in the CMB or the distribution of massive objects in the late universe (the mass function). To make a useful comparison between different statistics, we must extract the parameter measuring the strength of the interaction. Absent a complete theoretical prediction for all the moments, this means in practice that we expand about a Gaussian in the higher moments. Any valid truncation of this expansion depends on the structure of the moments, so we must be cautious about interpreting the results when a particular expansion has been assumed. 

As a case in point, it is tempting to try to use rare objects, the most massive clusters, to constrain or find evidence for non-Gaussianity 
\cite{Jee:2009nr, Brodwin:2010ig, Holz:2010ck, Cayon:2010mq,Enqvist:2010bg}.  Such measurements provide valuable information on scales different 
from CMB observations and, moreover, they are subject to a different set of systematics.  While there are many statistical issues in 
making such an analysis convincing \cite{Mortonson:2010mj, Hotchkiss:2011ms, Chongchitnan:2011eq, Hoyle:2011pj}, we will show here that we must also be very careful in interpreting what such objects imply for the physics of the primordial fluctuations. On the other hand, given good enough 
observations, the structure of the cumulants could be extracted from Large Scale Structure (LSS) measurements, yielding a remarkable window 
into the interaction structure of the underlying field theory. 

This paper is organized as follows.  In section \ref{sec:shift} we review single field inflation with an approximate shift symmetry, using
effective field theory to classify the possible non-Gaussian effects.  Section \ref{sec:structure} discusses the scaling of the correlation
functions that is associated with each type of non-Gaussianity.  Section \ref{sec:resc} considers another example of what we will call the feeder mechanism.  Next, Section \ref{sec:observational} considers the observational implications and discusses the possibility to disentangle various classes of microscopic interactions. We present generalized expressions for the halo mass function and for the Minkowski functionals, suitable for scenarios with non-hierarchical scaling.  Finally, in section \ref{sec:conclusions}, we conclude.  Appendix A reviews the classification of inflationary models, Appendix B provides a detailed computation of the non-Gaussian correlation functions from inverse decay and Appendix C provides some details on the computation of non-Gaussian correlators from rescattering.

\section{Inflation with a Shift Symmetry}
\label{sec:shift}

The standard ``vanilla'' inflationary models contain only a single scalar field with Lagrangian
\begin{equation}
\label{vanilla}
  \mathcal{L} = -\frac{1}{2}(\partial\varphi)^2 - V(\varphi) + \cdots
\end{equation}
where $\cdots$ denotes other fields, including the Standard Model of particle physics.  In order to support a long quasi de Sitter phase with $|\dot{H}| \ll H^2$ (here $H = \dot{a}/a$ is the Hubble scale and $a(t)$ is the scale factor) we require a flat scalar potential, quantified by the slow roll parameters
\begin{equation}
\label{SR}
  \epsilon = \frac{M_p^2}{2}\left(\frac{V'}{V}\right)^2 \, , \hspace{5mm} \eta =M_p^2 \frac{V''}{V} \, .
\end{equation}
In addition to the classical condensate, $\phi(t) \equiv \langle \varphi(t,{\bf x}) \rangle$, quantum fluctuations are inevitably present due to the inflationary background:
\begin{equation}
\label{split}
  \varphi(t,{\bf x}) = \phi(t) + \delta\varphi(t,{\bf x}) \, .
\end{equation}
The (super-horizon) curvature perturbation is obtained by a gauge transformation
\begin{equation}
\label{convert}
  \zeta(t,{\bf x}) = -\frac{H}{\dot{\phi}} \, \delta\varphi(t,{\bf x}) \, .
\end{equation}
Thus, in this scenario the \emph{same} field supports the background evolution and also generates the observed primordial curvature fluctuations. The power spectrum, $\mathcal{P}_\zeta(k)$, is defined by
\begin{equation}
\label{calP}
  \langle \zeta^2({\bf x}) \rangle = \int \frac{dk}{k} \mathcal{P}_\zeta(k) \, .
\end{equation}
In the vanilla scenario we have $\mathcal{P}^{1/2}_\zeta(k) = \frac{H^2_\star}{2\pi|\dot{\phi}_\star|}$ where quantities on the right-hand-side are understood to be evaluated at horizon crossing, $k=aH$.  From measurements of the amplitude of fluctuations in the cosmic microwave background \cite{Komatsu:2010fb}, $\mathcal{P}_{\zeta}\approx2.42\times10^{-9}\left(\frac{k}{0.002{\rm Mpc}^{-1}}\right)^{n_s-1}$ with the spectral index $n_s-1\approx0.034$. 

In this simple framework we can identify two important scales: $H$ and $\sqrt{\epsilon} M_p$ (the scale appearing in the conversion (\ref{convert}) between $\delta\varphi$ and $\zeta$).  These scales characterize most of the interesting low energy physics associated with the scalar curvature fluctuations.  In addition, the Planck scale, $M_p \approx 2.4\cdot 10^{18}\, \mathrm{GeV}$, controls quantum gravity effects and provides a natural UV cut-off.  

In the simple model (\ref{vanilla}), the only interactions are gravitational or else arise through the slow roll potential.  In both cases, the interaction strength is controlled by slow roll parameters and non-Gaussianity is too small to be observable in the next decade or more \cite{Acquaviva:2002ud,Maldacena:2002vr,Seery:2005wm}.  Additional interactions, controlled by a new scale, $f \ll M_p$, will arise once we try to embed the theory (\ref{vanilla}) into a more complete framework.

\subsection{Effective Field Theory of the Inflaton}
\label{subsec:EFT}

The slow roll parameters (\ref{SR}) are notoriously sensitive to Ultra-Violet (UV) physics and $\epsilon,|\eta|\ll 1$ may require significant fine tuning.  An appealing way to achieve slow roll in a naturally is to invoke a softly broken shift symmetry
\begin{equation}
\label{shift-symm}
  \varphi \rightarrow \varphi + \mathrm{const} \, .
\end{equation} 
In the simplest models $\phi(t)$ can be considered as a clock and (\ref{shift-symm}) leads to the approximate time-translation invariance that one might regard, in the context of the full gravitational system, as the ``true'' symmetry of inflation.

From a field theoretical perspective, the most cogent way to implement the symmetry (\ref{shift-symm}) is to assume that the inflaton is a Pseudo-Nambu-Goldstone-Boson (PNGB) which results from the spontaneous breaking of some global symmetry in the matter sector, at the scale $f$. The low energy theory will be characterized by the Goldstone mode, $\varphi$, along with any other light fields which are present, for example gauge fields. The original implementation of this idea was Natural Inflation \cite{Freese:1990rb,Adams:1992bn,Freese:1994fp,Savage:2006tr}, and many variations on this theme have followed \cite{Kinney:1995cc,Kim:2004rp,ArkaniHamed:2003wu,Dimopoulos:2005ac,Easther:2005zr,McAllister:2008hb,Flauger:2009ab,Berg:2009tg,Anber:2009ua,Kaloper:2008fb,Kaloper:2011jz,Kallosh:2007cc,Kawasaki:2000yn,Misra:2007cq,Grimm:2007hs, Kim:2011je}, motivated in part by the fact that such models may naturally generate an observably large amplitude of primordial gravitational waves.\footnote{An observable primordial gravitational wave background requires that the inflaton moves a super-Planckian distance in field space during inflation \cite{Lyth:1996im}. Ensuring that the potential is flat over such large displacements seems unnatural, unless there is extra physics at work such as an underlying shift symmetry.}  Despite this significant body of work, it was only recently that the cosmological perturbations in slow roll models with an underlying shift symmetry were understood comprehensively \cite{Barnaby:2010vf}. 

Following the standard logic of effective field theory, we write down terms in the action which are consistent with the symmetry (\ref{shift-symm}):
\begin{equation}
\label{L0}
  \mathcal{L}_{0} = -\frac{1}{2}(\partial\varphi)^2 - \frac{1}{4 f}\varphi G^a\tilde{G}^a - \frac{\alpha}{4 f} \varphi F\tilde{F} + \sum_{n=1}^\infty c_n \frac{(\partial\varphi)^{2n+2}}{f^{4n}} + \cdots
\end{equation}
Here $F_{\mu\nu} = \partial_\mu A_\nu - \partial_\nu A_\mu$ is the field strength for some $U(1)$ gauge field $A_\mu$, while $G^a_{\mu\nu}$ is the field strength associated with some non-Abelian gauge group.  (The gauge fields are assumed to have standard kinetic terms which we omit for ease of presentation.)  

The Langrangian (\ref{L0}) should be understood as an effective theory below the scale $f$.  At higher energies, the global symmetry is restored and one must include additional degrees of freedom beyond the PNGB.  Generically, the symmetry breaking scale $f$ also controls the strength of interactions in (\ref{L0}); absent fine tuning (or the appearance of additional scales in the problem) we expect $c_n = \mathcal{O}(1)$ and $\alpha = \mathcal{O}(1)$.

Notice that the shift symmetry (\ref{shift-symm}) strongly constrains the allowed interactions, and so will have important implications for observable non-Gaussian signatures. Goldstone's theorem implies that $\varphi$  can only have derivative interactions.\footnote{See \cite{Burgess:1998ku} for a review.}  These include the self-interaction terms $(\partial\varphi)^n$ and also the pseudo-scalar couplings to gauge fields, $\varphi F\tilde{F}$ and $\varphi G^a\tilde{G}^a$, that we have already included in (\ref{L0}).\footnote{The quantity $F\tilde{F}$ is a total derivative so that $\varphi F\tilde{F}$ gives a derivative interaction $\partial_\mu \varphi K^\mu$ after integrating by parts.  Similarly, also, for the non-Abelian gauge field.}  In addition to these leading order terms, higher derivative couplings to gauge fields are also allowed, suppressed by more powers of $f$.  Moreover, couplings to fermions will also be generically present; these will play no role in our discussion.

To drive inflation, we must slightly break the shift symmetry (\ref{shift-symm}).  This may be done in a controlled way, so that the smallness of the slow roll parameters (\ref{SR}) is protected by the mildness of the symmetry breaking effects.  There are two approaches: 
\begin{enumerate}
\item {\bf Non-Perturbative Breaking:} Instanton configurations of the non-Abelian gauge field will generically break the shift symmetry, 
  even if it is exact at the classical level.  The coupling $\varphi G^a\tilde{G}^a$ leads to a potential of the form
  \begin{equation}
  \label{Vnp}
    V_{\mathrm{np}}(\varphi) = \Lambda^4 \cos\left(\frac{\varphi}{f}\right) + \cdots
  \end{equation}
  where $\Lambda$ is a nonperturbatively generated scale and $\cdots$ denotes higher harmonics.  Notice that a discrete subgroup of the original
  symmetry remains unbroken: $\varphi \rightarrow \varphi + 2 \pi f$. 
 
 \item {\bf Explicit breaking:} One may instead break the symmetry (\ref{shift-symm}) explicitly at tree level.  We parametrize such effects by a non-periodic potential $V_{\mathrm{ex}}(\varphi)$ which may, for example, have a power-law form $\mu^{4-p} \varphi^p$.  See \cite{Kaloper:2008fb,Kaloper:2011jz} for an interesting field theory construction.  Explicit breaking may also be motived from a string theoretic perspective, for example, by wrapping branes on suitable cycles \cite{McAllister:2008hb,Flauger:2009ab}.
\end{enumerate}

Putting everything together, including symmetry breaking terms, we are led to consider the following Lagrangian:
\begin{equation}
 \mathcal{L} = -\frac{1}{2}(\partial\varphi)^2 - V(\varphi) - \frac{\alpha}{4 f}\varphi F\tilde{F} + \sum_{n=1}^\infty c_n \frac{(\partial\varphi)^{2n+2}}{f^{4n}} + \cdots \label{Ltot}
\end{equation}
where the potential is
\begin{equation}
\label{V}
  V(\varphi) = V_{\mathrm{ex}}(\varphi) + \Lambda^4 \cos\left(\frac{\varphi}{f}\right) + \cdots
\end{equation}
Note that in this simple framework there is a single scale, $f$, controlling the interactions in (\ref{Ltot}).  The introduction of explicit symmetry breaking effects (via $V_{\mathrm{ex}}$) may introduce an additional scale.  However, explicit symmetry breaking effects are assumed to be parametrically small, in order to preserve the slow roll conditions (\ref{SR}).  Hence, we expect that any additional interactions induced by explicit symmetry breaking will be too weak to contribute an observable non-Gaussian signature.\footnote{See \cite{Barnaby:2011qe} for a proof of this statement.}

\subsection{Computational Control}
\label{subsec:control}

In the last subsection, we have argued that (\ref{Ltot}) provides a generic low energy effective description of a PNGB inflaton, neglecting gravity.  This theory is characterized by a single scale, $f$.  However once we couple (\ref{Ltot}) to gravity, new scales arise in the problem associated with the curvature (Hubble) scale, its rate of evolution, and the Planck mass. In the inflationary context we should re-examine what it means to call the complete theory ``effective''.

We have in mind that $\varphi$ is the Goldstone mode associated with a global symmetry broken at the scale $f$.  In this case, we should presumably require $f\ll M_p$ since otherwise symmetry breaking would occur above the quantum gravity scale where Quantum Field Theory (QFT) is expected to become invalid.  Moreover, it is expected that quantum gravity effects (such as the virtual appearance of black holes) will explicitly break the original global symmetry \cite{ArkaniHamed:2003wu} and lead to corrections to (\ref{Ltot}) that are suppressed by powers of $f/M_p$.  Note that the condition $f\ll M_p$ seems also to be a requirement of string theory \cite{Banks:2003sx}, and is consistent with the ``gravity as the weakest force'' conjecture \cite{ArkaniHamed:2006dz}.

As discussed previously, we can identify two important scales -- $H$ and $\sqrt{\epsilon} M_p$ -- which characterize most of the physics associated with the curvature perturbations in the vanilla scenario. The non-gravitational sector (\ref{Ltot}) contains an additional scale, $f$, which arises naturally once we try to embed single-field inflation into a more complete UV framework, so it is helpful to organize our discussion using the ratio of the scale associated with field theory interactions, $f$, to the scales associated with gravity, $H$ and $\sqrt{\epsilon}M_p$.  We introduce the dimensionless quantities
\begin{equation}
\label{betas}
\beta_{M_p}\equiv\frac{\sqrt{\epsilon}M_p}{f} \, , \hspace{5mm} \beta_{H}\equiv\frac{H}{f} \, .
\end{equation}
These are of course not both independent of the amplitude of fluctuations, but they are useful to organize the physics.  

In order for the effective description (\ref{Ltot}) to make sense, we should require that inflation takes place \emph{below} the scale of the symmetry breaking that gives rise to the PNGB inflaton.  This implies that $\beta_H \ll 1$.  For larger values of $H$ we would need to include additional degrees of freedom, for example the massive field whose Vacuum Expectation Value (VEV) breaks the original global symmetry. 

Control over the low-energy description (\ref{Ltot}) often also requires that the higher dimension operators fall off with increasing 
powers of some small parameter.  Otherwise, we would need to re-sum an infinite series of such terms, the explicit form of which is not, in 
general, known.  The condition $\beta_H \ll 1$, which we have already discussed, helps to control higher derivative interactions involving the fluctuations, $(\partial \delta\varphi)^n$.  This is easily seen by estimating the derivatives as $\frac{\partial}{f} \sim \frac{H}{f} \sim \beta_H$. 
A more stringent constraint on model parameters comes from requiring that we can truncate the derivative expansion in (\ref{Ltot}) when studying
the dynamics of the homogeneous background, $\phi(t)$.  In this case one has
\begin{equation}
\label{derivative}
  \frac{\dot{\phi}^2}{f^4} \sim \left(\frac{\sqrt{\epsilon}M_p}{f}\right)^2 \left(\frac{H}{f}\right)^2 \sim \beta_{M_p}^2 \beta_{H}^2 \ll 1 .
\end{equation}
The higher derivative terms in (\ref{Ltot}) modify the sound speed, $c_s$, of the inflaton fluctuations.  Note that the condition (\ref{derivative}) implies that the sound speed is not too different from unity:
\be
  c_s^2\approx1-4\left(\frac{\sqrt{\epsilon}M_p}{f}\right)^2\left(\frac{H}{f}\right)^2=1-4\beta_{MP}^2\beta_H^2 \, \approx 1.
\ee
It is possible that some underlying symmetry principle strongly constrains the structure of derivative self-couplings, making it possible to re-sum the series and obtain $c_s \ll 1$ in a controllable way.  DBI inflation \cite{Alishahiha:2004eh} provides an example of such a theory, where the UV symmetry is associated with higher-dimensional Lorentz invariance \cite{deRham:2010eu}. One might also invoke Galileon symmetry \cite{Burrage:2010cu}.  

In summary, we generically expect that the low-energy effective description (\ref{Ltot}) will be under computational control provided 
$\beta_H \ll 1$ and $\beta_H \beta_{M_p} \ll 1$.\footnote{The latter condition can be relaxed in cases where we have more information
about the UV structure of the theory.}  Notice that there is a priori \emph{no} restriction on the size of $\beta_{M_p}$ by itself, however, we do require $f\ll M_p$.  In most interesting cases, the conditions for the validity of the effective field theory (\ref{Ltot}) can be summarized as:
\begin{equation}
\label{bnd}
  H \ll f \ll M_p \, .
\end{equation}

\subsection{Mechanisms for Non-Gaussianity}
\label{subsec:shiftNG}

Although our low energy effective description (\ref{Ltot}) is simple, it nevertheless allows for a rich array of non-Gaussian signatures.  Note that the \emph{same} shift symmetry (\ref{shift-symm}) which protects the flatness of the potential (\ref{V}), also strongly constrains the interactions of the inflaton.  Hence, this symmetry has important implications for non-Gaussianity.  There are three kinds of interactions which may be distinguished, each leading to a unique non-Gaussian effect:  

\begin{enumerate}

 \item {\bf Derivative Self-Interactions:} The higher derivative self-interactions $(\partial\varphi)^n$ lead to an equilateral bispectrum with nonlinearity parameter $f_{NL} \sim c_s^{-2}$ \cite{Chen:2006nt}.  An observationally interesting signal is possible for $c_s \ll 1$.

 \item {\bf Resonant Effects:} A different source of non-Gaussianity arises due to the periodic contribution to the potential (\ref{V}), which has its origin in nonperturbative symmetry breaking effects.  The self-interactions encoded in this term give rise to a novel oscillatory signal in the bispectrum \cite{Chen:2008wn,Flauger:2010ja,Leblond:2010yq,Hannestad:2009yx}.

 \item {\bf Pseudo-Scalar Couplings (Inverse Decay, Feeder field):} The pseudo-scalar interaction with Abelian gauge fields, $\varphi F\tilde{F}$, leads to a nonperturbative production of gauge quanta which, in turn, source non-Gaussian inflaton fluctuations via inverse decay \cite{Barnaby:2010vf,Barnaby:2011vw}. This mechanism was first understood in \cite{Barnaby:2010vf} and is reviewed in Appendix B.  Non-Gaussianity is observable when $f\ll M_p$, which seems required in any case for the consistency of the effective field theory description.  This scenario is a realization what we will call the ``feeder" mechanism, discussed in detail below.
\end{enumerate}

In all cases the interactions which source non-Gaussianity occur within (or close to) the horizon.  Moreover, in all cases there is no significant iso-curvature perturbation on large scales, nor super-horizon evolution of $\zeta$.

\subsection{A Note About ``Single Field" Inflation}

Before moving on, we pause to define how we distinguish between single field and multi-field models. Obviously, it is not useful to define a 
model as ``single field'' when only one degree of freedom appears in the Lagrangian: any model that incorporates realistic particle physics 
must necessarily include \emph{many} spectator fields which play no role in the inflationary dynamics. A more cogent definition is to 
classify a model as ``single field'' whenever there is only one field driving the background inflationary dynamics and that \emph{same} field 
is also responsible for generating the super-horizon curvature perturbation, $\zeta$. Throughout this paper it is understood that the expression ``single field'' has this meaning. Appendix A elaborates on this definition, establishing a taxonomy of inflationary models that we find useful to organize the relevant physics for the background and fluctuations. This is also relevant for understanding where the ``feeder'' mechanism sits in relation to the ``single field" effective field theory of the fluctuations \cite{Cheung:2007st}.

Notice that there is no obstruction to allowing a ``single field'' inflaton to couple directly to matter fields, as would seem to be a pre-requisite for successful reheating \cite{Kofman:1997yn,Barnaby:2004gg,Barnaby:2009wr,Braden:2010wd}.  (Here we use the word ``matter'' to denote any particles that are not directly involved in driving the background inflationary dynamics.)  Hence, the model (\ref{Ltot}) is classified as single field, despite the fact that we have explicitly included couplings to gauge fields.

\subsection{The Feeder Mechanism: Unexpected Results in Single Field Inflation}
\label{subsec:feeder}

The pseudo-scalar coupling discussed in subsection \ref{subsec:shiftNG} is actually an example of a larger class of scenarios that are ``single field'' according to our definition, but which exhibit novel and unexpected phenomenology \cite{Barnaby:2010vf,Barnaby:2011vw,Barnaby:2009mc,Barnaby:2009dd,Barnaby:2010ke,Barnaby:2010sq}. In particular, these models may give rise to observably large non-Gaussianity without recourse to any of the phenomena that add a layer of explanation on top of the inflationary shift-symmetry, including effects such as small sound speed \cite{Chen:2006nt}, higher derivatives \cite{Barnaby:2008fk,Barnaby:2007yb}, special initial conditions \cite{Chen:2006nt,Holman:2007na}, potentials with sharp features \cite{Chen:2006xjb,Chen:2006nt}, dissipative effects \cite{Green:2009ds}, fine-tuned inflationary trajectories, post-inflationary effects (such as preheating \cite{Barnaby:2006cq,Barnaby:2006km,Bond:2009xx}), etc.

The key mechanism that underlies this novel class of single-field inflation models works as follows.  From an effective field theory perspective, we generically expect direct couplings between the inflaton and one or more matter fields, as would seem to be required for successful reheating.  In some cases, the coupling to the time-dependent condensate can excite quanta of the matter field.  These produced matter particles may, in turn, source inflaton fluctuations via the \emph{same} coupling.  Inflaton fluctuations that are generated in this way tend to be highly non-Gaussian, and are complementary to the usual (nearly) Gaussian quantum fluctuations from the vacuum.  When this effect is observationally significant, we refer to the matter field as a \emph{feeder}, since its fluctuations grow in the environment of the quasi de Sitter background and, in turn, feed the inflaton perturbations, helping them to grow large, non-Gaussian and strong.

To date, several models have been explored which exhibit the feeder mechanism 
\cite{Barnaby:2010vf,Barnaby:2011vw,Barnaby:2009mc,Barnaby:2009dd,Barnaby:2010ke,Barnaby:2010sq,Cook:2011hg}.\footnote{See Appendix B for a 
review of the model proposed in \cite{Barnaby:2010vf}.}  These models differ in the details of the particle production (either tachyonic 
instability \cite{Barnaby:2010vf} or an instantaneous violation of adiabaticity \cite{Barnaby:2009mc}) and also the feed-back mechanism 
(inverse decay \cite{Barnaby:2010vf} or rescattering \cite{Barnaby:2009mc}).  However, the basic physics is very much analogous.  We expect 
that many other examples can be found. In \cite{Cook:2011hg,Senatore:2011sp} the production of gravitational waves via the feeder mechanism was studied.

In all known examples, the ``feeding'' dynamics take place entirely on scales inside or comparable to the horizon and the bispectrum is closer to equilateral than local.  There is no significant iso-curvature perturbation on super-horizon scales.  For $k \ll a H$ the curvature perturbation depends only on the inflaton field and is frozen, as in the vanilla scenario, so the feeder models really are ``single field'' in the conventional sense, both at the level of the homogeneous background dynamics and also at the level of the perturbations. 

\section{Structure of Correlation Functions}
\label{sec:structure}

In subsection \ref{subsec:shiftNG} we discussed three mechanisms -- sound speed, resonance and inverse decay -- which can give rise to 
observable non-Gaussian effects in the context of single field inflation with an underlying shift symmetry.  In this section, we will study 
the structure of correlation functions from each of these mechanisms, showing how this pattern of moments encodes information about the 
underlying microphysics.  In section \ref{sec:observational}, we will discuss how this information might be extracted from an observational perspective.

\subsection{The Dimensionless Moments}

To compare models we will work with the amplitude of the dimensionless moments, strictly defined from the equilateral limit of scale-invariant momentum space $n$-point functions.  We write the $n$-th cumulant\footnote{Here we have omitted any smoothing, which is irrelevant since our discussion will focus only on the scaling of the integrand, for the time being.  In section \ref{sec:observational} our definition of the moments will be made rigorous.} (the connected part of the real space $n$-point function) as
\begin{equation}
  \langle \zeta^n({\bf x}) \rangle_c \equiv \int \frac{d^3k_1}{(2\pi)^3} \frac{d^3k_2}{(2\pi)^3}\cdots \frac{d^3k_n}{(2\pi)^3} 
  \left[  (2\pi)^3 P_{n,\zeta}(k_i) \delta^{(3)}\left(\sum_{i=1}^n {\bf k_i}\right)  \right] \, .
\end{equation}
When the underlying fluctuations are exactly scale invariant, the polyspectra scale as $P_{n,\zeta} \sim k^{-3(n-1)}$.  The dimensionless 
moments, $\mathcal{M}_n$, are defined by
\be
\label{Mndef}
\mathcal{M}_n\equiv \frac{k^{3(n-1)}P_{n,\zeta}(k,k,\dots,k)}{\left[k^3 P_{2,\zeta}(k)\right]^{n/2}} 
  =  \frac{k^{3(n-1)}P_{n,\zeta}(k,k,\dots,k)}{\left(2\pi^2\mathcal{P}_{\zeta}\right)^{n/2}} \, ,
\ee
where $\mathcal{P}_\zeta$ is defined by (\ref{calP}) and we have taken the equilateral limit $|\vec{k}_1|=|\vec{k}_2|=\dots=|\vec{k}_n|=k$. For a Gaussian distribution $\mathcal{M}_{n\geq 3} = 0$, so these provide a measure of non-Gaussianity.

Note that because the $\mathcal{M}_n$ are dimensionless, we can use them as the dimensionless moments for either the primordial curvature $\zeta$ or for the matter era Bardeen potential, $\Phi=\frac{3}{5}\zeta$ by replacing $P_{n,\zeta}\rightarrow P_{n,\Phi}$ and $\mathcal{P}_{\zeta}\rightarrow\Delta^2_{\Phi}$. Throughout, we will ignore any scale-dependence (including $n_s\neq1$) that will make $\mathcal{M}_n$ a function of scale.  Such scale dependence is typically mild, although it may still be relevant for understanding the perturbation theory if it is stronger at higher $n$ (eg, see \cite{Leblond:2008gg}) and is likely to be observable if the non-Gaussianity is not too small \cite{LoVerde:2007ri, Sefusatti:2009xu, Shandera:2010ei}.  Note that, for resonant models, this convention amounts to ignoring the oscillatory functions that modulate the amplitude.  In Section \ref{sec:observational} we will move the discussion to real space, and then these approximations can be avoided, but a smoothing scale is typically introduced which also changes the scaling slightly. Since the $\mathcal{M}_n$ defined above establish the scaling of the coefficient of some momentum integrals even in the real space case, we will often schematically write $\mathcal{M}_n$ as $\frac{\langle\zeta^n\rangle_c}{\langle\zeta^2\rangle^{n/2}}$ without specifying the argument of $\zeta$.

In this paper we are interested in how the pattern of dimensionless moments encodes the physics of the underlying inflaton interactions.
A useful diagnostic of the pattern of moments is the ratio
\begin{equation}
\label{ratio}
  \frac{\mathcal{M}_{n+1}}{\mathcal{M}_n}
\end{equation}
We will focus on the parametric dependence of this ratio on the small parameter that controls the strength of interactions, showing that
very different patterns arise for self-interactions as compared to interactions with other sectors.  The ratio (\ref{ratio})
will also depend on numerical coefficients which may grow with $n$ due to combinatorics.  Such model-dependent characteristics are, of
course, different from the parametric pattern of moments and must be considered on a case-by-case basis.  In practice, observations of rare 
objects are only sensitive to relatively low order moments (certainly $n < 10$ is sufficient), so we do not expect combinatorics to modify 
our key results in any significant way (see section \ref{sec:observational}).  We leave a more detailed analysis to future works.

We will call a set of moments $\{\mathcal{M}_n\}$ {\it ordered} if $\mathcal{M}_{n+1}<\mathcal{M}_n$ for all $n$. Often, higher order moments fall off with increasing powers of some small parameter and so are naturally ordered. In this case we can develop a systematic expansion of the Probability Distribution Function (PDF) around the Gaussian; see section \ref{sec:observational}.  Moreover, non-Gaussian corrections to the PDF are controlled by the smallness of the dimensionless moments.

\subsection{Self-Interactions and Hierarchical Scaling}
\label{subsec:selfint}

Our effective field theory description (\ref{Ltot}) contains two distinct classes of interactions: self-interactions (leading to resonance or sound speed effects) and couplings to other sectors (leading to inverse decay effects).  In the simple scenario we consider, both kinds of effects are controlled by a single scale, $f$.  Shortly, we will verify explicitly that the the self-interaction terms in (\ref{Ltot}) lead to a specific pattern of moments referred to as \emph{hierarchical}.  To shed light on the underlying physics, it is interesting to first consider this point from a somewhat more general perspective.

Suppose the fundamental Lagrangian contains only interactions that couple $\zeta$ to itself and, moreover, there is a single scale defining the interaction strength, $\mathcal{I}$ in appropriate units. Heuristically, we expect the ratio of the $(n+1)$-function to the $n$-function to scale as
\begin{equation}
\label{naive-ratio}
  \frac{\langle \zeta^{n+1}\rangle_c }{\langle \zeta^n \rangle_c} \propto \mathcal{I} \cdot2\pi^2\mathcal{P}_\zeta \, .
\end{equation}
The factor of $\mathcal{I}$ is obvious: higher correlation functions involve either higher order interaction terms or additional vertices, which necessarily carry more powers of the interaction strength.  The factor of $\mathcal{P}_\zeta$ arises because the $(n+1)$-th correlator has an additional external line, as compared to the $n$-th correlator.  To connect this external line to the rest of the diagram requires an additional propagator, which scales as $\zeta^2 \sim \mathcal{P}_\zeta$.  We define our normalization so that the $2$-point function gives one power of $\mathcal{P}_\zeta$ and the $3$-point defines one power of the interaction strength.  Putting these together, we expect
\be
\label{naive-scale}
\langle\zeta^n\rangle\propto\left(\mathcal{I}\right)^{n-2}\left(2\pi^2\mathcal{P}_{\zeta}\right)^{n-1} \, .
\ee
Then the dimensionless moments (\ref{Mndef}) scale as
\be
\label{eq:hierarchical}
\mathcal{M}_n\propto\left(\mathcal{I}^22\pi^2\mathcal{P}_{\zeta}\right)^{\frac{n-2}{2}} \, .
\ee
This is the scaling commonly referred to in the literature as \emph{hierarchical}, and the moments are \emph{ordered} provided
\begin{equation}
\label{Ihier}
  \mathcal{I} \cdot \sqrt{2}\pi\mathcal{P}_\zeta^{1/2} \ll 1 \, .
\end{equation}
In many examples, this same condition also controls the loop expansion.

If there are additional scales in the problem, or fine-tuning, it may be possible to obtain non-hierarchical moments from self-interactions.  Nevertheless, as we will see explicitly, the heuristic argument which we have presented does apply to the most widely-studied non-Gaussian mechanisms.

\subsection{The Local Ansatz}
\label{subsec:local}

In order to illustrate the statements made in the last subsection, let us consider the so-called local ansatz for the curvature perturbation:
\begin{equation}
\label{local}
 \zeta = \zeta_g + \frac{3}{5} f_{NL}\left[\zeta_g^2 - \langle \zeta_g^2 \rangle \right] \, ,
\end{equation}
where $\zeta_g$ is a Gaussian random field.  This expression provides a useful phenomenological description of non-Gaussianity and may also arise in multi-field or curvaton models.  The cumulants scale as $\langle\zeta^n\rangle_c \sim \left(f_{NL}\right)^n \langle\zeta_g^2\rangle^{n-1}$ and hence the dimensionless moments have the structure
\begin{equation}
  \mathcal{M}_n^{\mathrm{local}} = A_n \left[\left( \frac{3}{5} f_{NL} \right)^2 2\pi^2\mathcal{P}_\zeta \right]^{\frac{n-2}{2}} \, ,
\end{equation}
where $A_n$ is some numerical coefficient from combinatorics. The local ansatz reproduces (\ref{eq:hierarchical}) with $\mathcal{I} \sim f_{NL}$.  The moments are ordered when (\ref{Ihier}) is satisfied.  Equivalently, we can write $f_{NL} \langle \zeta_g^2 \rangle^{1/2} \ll 1$.  This shows that the naive condition for the non-Gaussian part of (\ref{local}) to give a subdominant contribution to the two-point function also ensures that the moments are ordered.

In general, scenarios with hierarchical scaling need not give a non-Gaussianity of the type (\ref{local}).  Rather, the bispectrum may have a shape which is very different from local. Very different bispectra are often compared by defining an effective nonlinearity parameter as
\begin{equation}
\label{fNLeff}
  f_{NL}^{\mathrm{eff}} \sim \frac{5}{3}\frac{\langle \zeta^3 \rangle}{\langle \zeta^2 \rangle^2} \, ,
\end{equation}
where the bispectrum has been evaluated in the equilateral limit and the factor of $5/3$ makes this convention consistent with the usual local ansatz. For some models, this definition is consistent with $\mathcal{I} \sim f_{NL}^{\mathrm{eff}}$ so that our generic result, equation (\ref{eq:hierarchical}), is brought into the familiar form
\begin{equation}
\label{MfNLeff}
 \mathcal{M}_n \propto \left[(f_{NL}^{\mathrm{eff}})^2 2\pi^2\mathcal{P}_\zeta \right]^{\frac{n-2}{2}} \, .
\end{equation}
However, we will see in two examples below (the inflaton/curvaton scenario and resonant non-Gaussianity) the definition of $f_{NL}^{\mathrm{eff}}$ directly from the bispectrum does not always correctly capture the interaction strength, while the ratio of consecutive dimensionless moments (\ref{ratio}) does.  It is worth emphasizing that the role of $f_{NL}^{\mathrm{eff}}$ in our discussion is to provide a simple measure of the parameter which controls the strength of interactions.  The full bispectrum contains information beyond the amplitude -- the shape and running of non-Gaussianity are both important discriminating tools -- but these are not the subject of our attention here.  Indeed,  we will ultimately be interested in observables such as the halo mass function and Minkowski functionals, which are relatively insensitive to the detailed shape of non-Gaussianity.

\subsection{Inflaton/Curvaton non-Gaussianity}

An interesting variation on the local model comes when two fields contribute to the curvature, one of which has local-type non-Gaussianity. In other words, an Antatz of the type:
\be
\label{InfCurv}
\zeta(x)=\phi(x)+\sigma(x)+\frac{3}{5}\tilde{f}_{NL}\sigma(x)^2
\ee
Defining the ratio of power in the two-point from the two fields as
\be
\xi^2=\frac{\mathcal{P}_{\sigma}}{\mathcal{P}_{\sigma}+\mathcal{P}_{\phi}}=\frac{\mathcal{P}_{\sigma}}{\mathcal{P}_{\zeta}}\;.
\ee
we see that the leading terms in the higher moments go like
\be
\mathcal{M}_n\propto\xi^2\left(\frac{3}{5}\tilde{f}_{NL}\xi^2\right)^{n-2}(2\pi^2\mathcal{P}_{\zeta})^{\frac{1}{2}(n-2)}\equiv\xi^2\left(\mathcal{I}^22\pi^2\mathcal{P}_{\zeta}\right)^{\frac{n-2}{2}}\;.
\ee
These moments have the hierarchical scaling, although with an extra parametric dependence on $\xi^2$ in each of the dimensionless moments so that defining $f_{NL}^{{\rm eff}}$ as in Eq.(\ref{fNLeff}) above would be misleading as a measure of the interaction strength. Notice that if the real space expansion Eq.(\ref{InfCurv}) had no term linear in $\sigma$ above this would become the ``un-Gaussiton" which scales like the feeder mechanism we discuss beginning in subsection \ref{subsec:differentscaling} below.

\subsection{Sound Speed Effects}
\label{subsec:cs}

Let us now consider a less trivial example: non-Gaussianity generated by the derivative self-interactions in (\ref{Ltot}).  Models of this type have been well-studied in the literature and it is known that the strength of interactions is controlled by the sound speed, $c_s$, and that large non-Gaussianity is possible when $c_s \ll 1$.  (In terms of the scale $f$ suppressing the derivative interactions, small $c_s$ indicates $f$ much closer to $H$ than to $\sqrt{\epsilon}M_p$ \cite{Shandera:2008ai,Baumann:2011su}.) The dimensionless moments for generic derivative self-interactions were estimated in \cite{Leblond:2008gg, Shandera:2008ai}
\be
\label{MomentsSoundSpeed}
\mathcal{M}_n^{c_s} = A_n \left(c_s^{-4} 2\pi^2\mathcal{P}_\zeta \right)^{\frac{n-2}{2}} \, ,
\ee
where $A_n$ is independent of model parameters.  This is exactly the form (\ref{eq:hierarchical}) with 
$\mathcal{I} \sim f_{NL}^{\mathrm{eff}} \sim c_s^{-2}$, consistent with the expectation that the sound speed 
should be associated with the strength of interactions.

If we consider an action with an arbitrary function of powers of the field and its first derivative, we would find additional parameters that describe the amplitude of other non-Gaussianity terms at each order $n$. However, each parameter generates a family of higher moments with the hierarchical scaling. Explicit examples can be found in \cite{Huang:2006eh}. If the structure of higher derivative terms is altered by imposing an additional symmetry, the scaling may also change \cite{Creminelli:2010qf}.

\subsection{Resonance Effects}
\label{subsec:resonance}

The periodic contribution to the potential (\ref{V}), which arises due to nonperturbative symmetry breaking effects, provides another interesting source of non-Gaussianity.  Expanding the field in terms of fluctuations as (\ref{split}) we have interaction terms of the form
\begin{equation}
\label{resint}
  \mathcal{L}_n = -\frac{\Lambda^4}{f^n} \frac{\Psi_n(t)}{n!} \left(\delta\varphi\right)^n
\end{equation}
where the oscillatory time dependence of the effective coupling is encoded by
\begin{equation}
 \Psi_n(t) \equiv \left\{ \begin{array}{ll}
         (-1)^{n/2} \cos\left(\frac{\phi(t)}{f}\right) & \hspace{5mm}\mbox{$n$ even};\\
         (-1)^{(n+1)/2} \sin\left(\frac{\phi(t)}{f}\right) & \hspace{5mm} \mbox{$n$ odd}.\end{array} \right.
\end{equation}
Non-Gaussianity arises due to a resonance between the oscillation frequency of the effective coupling and the wavenumber $k$ of a given mode.  Resonance occurs inside the horizon, leading to a bispectrum that is closer to equilateral than local.

The higher order correlation functions due to the interaction (\ref{resint}) were recently computed \cite{Leblond:2010yq}.  At leading order the result is
\be
\label{MomentsResonant}
\mathcal{M}_n^{res.}=A_n \, \beta_{M_p}^{-\frac{1}{2}}\:\left(\beta_{M_p}^42\pi^2\mathcal{P}_{\zeta}\right)^{\frac{n-2}{2}}\; ,
\ee
where $A_n$ are dimensionless numbers and $\beta_{M_p}$ was defined in (\ref{betas}).\footnote{The quantity $\beta_{M_p} \equiv \sqrt{\epsilon} M_p / f$ was called $\alpha$ in reference \cite{Leblond:2010yq}.}  Hence, we recover the hierarchical structure (\ref{eq:hierarchical}) with $\mathcal{I}=\beta_{M_p}^2=(\sqrt{\epsilon}M_p/f)^2$.

It is interesting, also, to compare (\ref{MomentsResonant}) with (\ref{MomentsSoundSpeed}).  From Eq.(\ref{MomentsResonant}), it looks like $\beta_{M_p}^2$ plays the role of $1/c_s^2$ in the small sound speed models. However, there is an extra dependence on the parameter $\beta_{M_p}$ in the coefficient of each $\mathcal{M}_n$, so that the amplitude of the three point is given parametrically by $\beta_{M_p}^{3/2}$ instead of $\beta_{M_p}^2$. That is,
\be
\Expect{\zeta^3}\propto\beta^{3/2}_{M_p}\left(2\pi^2\mathcal{P}_{\zeta}\right)^2\;.
\ee
As with the inflaton/curvaton example, this highlights the fact that amplitude of the third moment less to distinguish models than information about the structure of the moments and their parametric dependence on the scales in the problem.\footnote{The full bispectrum contains more information than the skewness; the shape and running of non-Gaussianity are also important discriminators.  However, for the kinds of probes that we will ultimately be interested in -- the abundance of rare objects and morphological indicators -- it is the integrated  moments that are most relevant.  More on this later.} To check or uncover the pattern in Eq.(\ref{MomentsResonant}) one would need at least measurements of the three and four point amplitudes.
%As with the inflaton/curvaton example, this highlights the fact that the amplitude of the three point alone does less to distinguish models than information about the structure of the moments and their parametric dependence on the scales in the problem. To check or uncover the pattern in Eq.(\ref{MomentsResonant}) one would need at least measurements of the three and four point amplitudes.

Note that our discussion of resonance effects has focused on Fourier space amplitudes.  Here we characterize the size of non-Gaussian moments by the amplitude of their oscillatory polyspectra.  Care may be required to translate this particular discussion into real space where integration of the oscillatory functions over $k$ can lead to cancellations.

A final interesting point about the resonant non-Gaussianity is that at high enough order $n$, the correlation functions may be dominated by a contribution that scales slightly differently, with one less power of $\beta_{M_p}$. When $\beta_{M_p}$ is a large number, this improves the convergence of the series of moments \cite{Leblond:2010yq}.

\subsection{Inverse Decay Effects}
\label{subsec:inversedecay}

So far, we have seen that the self-interactions in (\ref{Ltot}) give rise to a hierarchical structure of correlation functions where the dimensionless moments fall off according to Eq.(\ref{eq:hierarchical}).  This exhausts two of the three mechanisms which may generate observable non-Gaussianity in the context of single field inflation with an underlying shift symmetry (resonance and sound speed effects).  We now explore the structure of correlation functions associated with the third mechanism: \emph{inverse decay effects}.  We will see that a dramatically different kind of scaling behaviour arises in this case.

Inverse decay effects generate non-Gaussian correlations in a way that is fundamentally different from the kind of self-interaction vertices that were discussed in subsection \ref{subsec:selfint}.  Hence, we do not expect that equation (\ref{eq:hierarchical}) will hold.  The key physics of inverse decay was first understood in \cite{Barnaby:2010vf} and is reviewed in Appendix B where we also compute the parametric dependence of \emph{all} $n$-point functions.  Let us briefly re-capitulate the key aspects in order to provide a heuristic estimate of the cumulants.

We are interested in the Lagrangian (\ref{Ltot}), assuming that resonance and sound speed effects are negligible.  The calculation proceeds in two steps.  First, consider the coupling of the gauge field to the time-dependent condensate
\begin{equation}
  \mathcal{L} \supset -\frac{1}{4}F^2 - \frac{\alpha \phi(t)}{4f} F\tilde{F} +\cdots
\end{equation}
For $\phi = \mathrm{const}$ the second term is a total derivative which has no impact on the classical equation of motion.  However, with 
$\dot{\phi}\not= 0$ this term breaks the conformal invariance of the gauge field action at a scale $M_\star \sim \frac{\alpha\dot{\phi}}{f}$.
This leads to a tachyonic production of gauge fluctuations with $\delta A_\mu \sim H e^{\pi \xi}$.  Here 
$\xi \equiv \frac{\alpha \dot{\phi}}{2 H f} \sim \frac{M_\star }{H}$ quantifies the scale associated with the breaking of 
conformal invariance as compared to the Hubble scale.

The next step in the calculation is the feed-back of the produced gauge field fluctuations into the inflaton perturbation.  The relevant terms in the Lagrangian are
\begin{equation}
 \mathcal{L} \supset -\frac{1}{2}(\partial\delta\varphi)^2 - \frac{m^2}{2}(\delta\varphi)^2 - \frac{\alpha}{4 f} \delta\varphi F\tilde{F} + \cdots
\end{equation}
The last term mediates inverse decay processes where two of the produced gauge fluctuations generate an inflaton mode $\delta A_\mu + \delta A_\mu \rightarrow \delta\varphi$.  Clearly, this term has a very different structure from the self-interaction vertex considered in subsection \ref{subsec:selfint}.  The amplitude of this effect can be naively estimated from the equation of motion
\begin{equation}
\label{ideqn}
  \left[\partial_t^2 + 3 H \partial_t - \frac{\nabla^2}{a^2} + m^2 \right] \delta\varphi = \frac{\alpha}{4 f} F\tilde{F} \, .
\end{equation}
The relevant dynamics takes place near horizon crossing, so we estimate the derivatives as $\partial \sim H$.  This, along with our previous estimate $\delta A_\mu \sim H e^{\pi \xi}$, suggests $\delta \varphi_{\mathrm{inv.dec}} \sim \frac{\alpha H^2}{f} e^{2\pi \xi}$.  Using $\zeta \sim -\frac{H}{\dot{\phi}}\delta\varphi$ we are led to expect $\langle \zeta^n \rangle \propto \left(\frac{H^2}{\dot{\phi}} \frac{\alpha H}{f} e^{2\pi\xi}\right)^n$.

In Appendix B we explicitly compute the parametric dependence of all non-Gaussian $n$-point functions generated by inverse decay.  The result
for the polyspectra is
\begin{equation}
\label{id_poly}
  P_{n,\zeta}(k) = \left( 0.0031 \frac{H^2}{2\pi|\dot{\phi}|}\frac{\alpha H}{f}\frac{e^{2\pi\xi}}{\xi^4} \right)^n \, \frac{A_n}{k^{3(n-1)}}
\end{equation}
See equation (\ref{inv.dec_cumulant}).  Here $A_n$ are dimensionless numbers, independent of model parameters.  By construction $A_2 = 1$ and it may be verified that $A_3 = \mathcal{O}(1)$ \cite{Barnaby:2010vf}.  We expect that the higher order coefficients are also order unity, although combinatorial factors may modify this expectation at very large $n$.  This is an interesting issue, but since most observables are not sensitive to extremely high order moments we will leave this matter to future investigation.  Equation (\ref{id_poly}) confirms the heuristic estimate in the last paragraph (up to a power-law dependence on $\xi$ which arises from carefully solving for the time-dependence of the gauge field fluctuations).

In addition to the inflaton fluctuations generated by inverse decay, we also have the usual quantum vacuum fluctuations from inflation, 
$\delta\varphi_{\mathrm{vac}} \sim H$ (the homogeneous part of the solution of (\ref{ideqn})).  These are nearly Gaussian and hence 
irrelevant for the $n=3$ and higher moments.  However, the $2$-point function gets important contributions from \emph{both} inverse decay and 
also vacuum fluctuations; see equation (\ref{pwr}) for the explicit result.  There are two limiting cases which must be distinguished, 
depending on which effect dominates
\begin{equation}
\label{2-pt-scale}
  \mathcal{P}_\zeta \approx \left\{  
  \begin{array}{ll}
    \frac{H^4}{(2\pi)^2\dot{\phi}^2}, & \hspace{5mm}\mbox{vacuum\hspace{2mm}dominated} \\
    \frac{H^4}{(2\pi)^2\dot{\phi}^2} \frac{1}{2\pi^2} \left( 0.0031 \frac{\alpha H}{f} \frac{e^{2\pi\xi}}{\xi^4} \right)^2, & \hspace{5mm}\mbox{inverse\hspace{2mm}decay\hspace{2mm}dominated}
  \end{array}
 \right.
\end{equation}
where the spectrum is defined as in (\ref{calP}).
The condition to be in the vacuum dominated regime is
\begin{equation}
\label{vacdomcond}
  0.0031 \frac{\alpha H}{f} \frac{e^{2\pi\xi}}{\xi^4} \ll \sqrt{2}\pi \hspace{5mm}\left[\mathrm{vacuum}\hspace{2mm}\mathrm{dominated}\right]
\end{equation}
which is evident from comparing the two limiting behaviours in (\ref{2-pt-scale}); see also (\ref{pwr}).  (Obviously, the inverse decay dominated regime corresponds to (\ref{vacdomcond}) with the inequality reversed.)

\subsection{A Different Kind of Scaling}
\label{subsec:differentscaling}

We are finally in a position to compute the structure of dimensionless moments from inverse decay effects.  It is straightforward to show that
\begin{equation}
\label{MomentsInvDec}
   \mathcal{M}_n^{\mathrm{inv.dec}} = \left\{  
  \begin{array}{ll}
    A_n\left( \frac{0.0031}{\sqrt{2}\pi} \frac{\alpha H}{f} \frac{e^{2\pi\xi}}{\xi^4}\right)^n, & \hspace{5mm}\mbox{vacuum\hspace{2mm}dominated} \\
    A_n, & \hspace{5mm}\mbox{inverse\hspace{2mm}decay\hspace{2mm}dominated}
  \end{array}
 \right.
\end{equation}
Let us consider each of these limiting behaviours separately.

\begin{enumerate}

\item {\bf Vacuum Dominated Regime:} Here the condition (\ref{vacdomcond}) is satisfied and the moments are easily seen to be 
ordered in the sense that $\mathcal{M}_{n+1}^{\mathrm{inv.dec}} \ll \mathcal{M}_{n}^{\mathrm{inv.dec}}$. However, the structure of correlation functions is very different from what would be obtained for self-interactions.  To see this, 
we introduce an effective nonlinearity parameter as in equation (\ref{fNLeff}).  The dimensionless moments can now be written as
\ba
 \mathcal{M}_n^{\mathrm{inv.dec}} &\propto& \mathcal{I}\,^n\\\nonumber
&\propto& \left(f_{NL}^{\mathrm{eff}} \mathcal{P}^{1/2}_\zeta\right)^{n/3} \hspace{5mm} \left[\mathrm{vacuum}\hspace{2mm}\mathrm{dominated}\right]
\label{idscales}
\ea
where in the first line we identify $\mathcal{I} \equiv \frac{0.0031}{\sqrt{2} \pi} \frac{\alpha H}{f} \frac{e^{2\pi\xi}}{\xi}$ as the natural
measure of the strength of interations in this scenario.  In the second line of (\ref{idscales}) we re-write the moments in a way that 
is appropriate for contrasting with small sound speed models, Eq.(\ref{MfNLeff}), and uses Eq.(\ref{fNLeff}) to define the effective 
non-linearity parameter. For a given value of $f_{NL}^{\mathrm{eff}}$ the moments from inverse decay fall off much more slowly than for 
self-interactions, so that even if both mechanisms are tuned to give a $3$-point function of the same size, the inverse decay scenario
will lead to a PDF that is intrinsically more non-Gaussian.  

\item {\bf Inverse Decay Dominated Regime:} In this case the dimensionless moments \emph{are not even ordered}!  This can be understood physically: when 
inverse decay effects dominate, $\zeta$ is bi-linear in the Gaussian field $\delta A_\mu$.  Hence, we do not expect the PDF to be close 
to Gaussian in any meaningful sense.  Remarkably, the computation which leads to (\ref{MomentsInvDec}) is still under control.  In Appendix B we discuss the constraints coming from calculability.  The most important consistency condition arises from ensuring that  backreaction effects are 
negligible.  This amounts to the constraint
\begin{equation}
  \sqrt{2}\pi \mathcal{I} = 0.0031 \frac{H^2}{2\pi |\dot{\phi}|} \frac{\alpha H}{f} \frac{e^{2\pi \xi}}{\xi^4} \ll  \mathcal{O}(1) \, .
\end{equation}
This constraint ensures that the $n$-point functions of $\zeta$ are small.  However, in the inverse decay dominated regime the small 
parameter cancels out of $\mathcal{M}_n$, leading to a non-ordered structure.

\end{enumerate}

We see, then, that inverse decay effects are remarkably different from the self interactions discussed in subsection \ref{subsec:selfint}.  
In the latter case, the \emph{same} small parameter which controls the hierarchical structure of moments is \emph{also} responsible for 
ensuring computational control.  In the inverse decay case, on the other hand, these two statements are effectively decoupled.  It is possible 
to have non-ordered moments in a computationally reliable regime.  We expect that similar results will hold also in other examples of
the feeder mechanism; more on this in the next section.

\subsection{Salvaging the Inverse Decay Dominated Regime}

The inverse decay dominated regime is theoretically interesting, since it provides a novel example of a non-hierarchical PDF which is 
computationally reliable.  However, this scenario gives $f_{NL}^{\mathrm{equil}} \gsim 10^3$, which is strongly disfavoured from an
observational perspective.  There are several ways to circumvent this difficulty.  One idea, put forward in \cite{Sorbo:2011rz}, is to 
consider $\mathcal{N} \gg 1$ $U(1)$ gauge fields, each coupled to the inflaton as in our model (\ref{Ltot}) with the \emph{same} coupling 
strength $\alpha/f$.  Because these gauge fields are statistically independent, the $n$-point correlation functions are multiplied by 
$\mathcal{N}$.  To hold fixed the normalization of the power spectrum we must require that the quantity $\mathcal{I}\propto\frac{H^2}{\dot{\phi}} \frac{\alpha H}{f} \frac{e^{2\pi\xi}}{\xi^{4}}$ scales as $\mathcal{N}^{-1/2}$.  Given this relation, the effective nonlinearity parameter must scale as $\mathcal{N}^{-1/2}$.  For $\xi \rightarrow \infty$ the nonlinearity parameter in the $\mathcal{N}=1$ model saturates to $f^{\mathrm{eff}}_{NL} \rightarrow 8400$.  Hence, in this regime we can write the result for the multi-field case as 
\begin{equation}
  f_{NL}^{\mathrm{eff}} = 8400 \, \mathcal{N}^{-1/2} \, ,
\end{equation}
so that $\mathcal{N} \sim 10^3$ is sufficient to render the model compatible with observational bounds on CMB scales.  The dimensionless
cumulants $\mathcal{M}_n$ scale as $\mathcal{N}^{1-n/2}$, giving
\begin{equation}
\label{salvage}
  \mathcal{M}_n^{\mathcal{N}} = \left( 1.2\cdot 10^{-4} f_{NL}^{\mathrm{eff}} \right)^{n-2} \mathcal{M}_n^{\mathcal{N}=1} \, .
\end{equation}
Remembering that $\mathcal{M}_n^{\mathcal{N}=1}$ are independent of model parameters, we see that the cumulants exhibit a scaling very 
similar to the standard hierarchical scaling, Eq. (\ref{MfNLeff}).  The key difference is that the small parameter $\mathcal{P}^{1/2}_\zeta f_{NL}^{\mathrm{eff}}$ has been replaced with $1.2\cdot 10^{-4} f_{NL}^{\mathrm{eff}}$.

\section{Another Example: Production of Massive Particles}
\label{sec:resc}

In section \ref{sec:structure} we studied the structure of correlation functions for the various scenarios that can lead to observable non-Gaussianity in the model (\ref{Ltot}).  We found that inverse decay effects lead to a radically different structure than self-interactions.  We expect that this novel pattern of moments will be shared by other examples of the feeder mechanism.  To verify this, let us now turn our attention to a model that was studied in detail in \cite{Barnaby:2009mc,Barnaby:2009dd,Barnaby:2010ke} (see \cite{Barnaby:2010sq} for a review).  This model is somewhat tangential to our original construction of a PBNG inflaton (\ref{Ltot}).  Nevertheless, we will see that many similarities arise.

Consider the theory
\begin{equation}
\label{Lresc}
 \mathcal{L} = -\frac{1}{2}(\partial\varphi)^2 - V(\varphi) 
  - \frac{1}{2}(\partial\chi)^2 - \frac{g^2}{2}(\varphi-\phi_0)^2\chi^2  \, ,
\end{equation}
where $\varphi$ is the inflaton and $\chi$ is a matter fields.  This theory provides a simple and explicit toy model to investigate the implications of particle production during inflation and shares many qualitative similarities with the inverse decay mechanism discussed previously.  Models of the type (\ref{Lresc}) are quite natural in the context of open string inflation and various extensions of this simple framework have been studied in \cite{Green:2009ds,Battefeld:2010sw,Battefeld:2011yj}. 

Let us briefly discuss the dynamics of the model (\ref{Lresc}).  At the moment when $\phi(t)=\phi_0$ the effective mass of $\chi$ vanishes, leading to an instantaneous violation of adiabaticity and associated particle production.  The variance of produced fluctuations is
\begin{equation}
 \langle \chi^2 \rangle = \frac{n_{\chi}}{g |\phi-\phi_0|} \, ,
\end{equation}
where the number density of produced quanta is $n_{\chi} \sim a^{-3} (g \dot{\phi})^{3/2}$ where the factor $a^{-3}$ reflects the usual volume
dilution of non-relativitic particles.  The dominant effect of particle production of the observable spectrum of curvature fluctuations 
arises because the produced, massive $\chi$ particles can rescatter off the condensate to generate bremsstrahlung radiation of 
long-wavelength $\delta\varphi$ fluctuations \cite{Barnaby:2009mc}.  This process is described by the equation
\begin{equation}
\label{resc_eqn}
 \left[ \partial_t^2 + 3 H \partial_t -\frac{\nabla^2}{a^2} + m^2 \right] \delta\varphi = - g^2 \left[\phi(t)-\phi_0\right] \chi^2 \, .
\end{equation}
(In Appendix C we discuss the theory of this equation in more detail.)  Rescattering leads to a localized, bump-like feature in the 
primordial power spectrum.  This may be described by the following simple fitting function \cite{Barnaby:2009dd}:
\begin{equation}
\label{Pfit}
  \mathcal{P}_\zeta(k) = A_s \left(\frac{k}{k_0}\right)^{n_s-1} 
    + A_b \left(\frac{\pi e}{3}\right)^{3/2}\left(\frac{k}{k_b}\right)^3 e^{-\frac{\pi}{2}\frac{k^2}{k_b^2}}
\end{equation}
where $k_0$ is the pivot and $k_b$ is the location of the feature.  The first term is associated with the usual vacuum fluctuations from 
inflation (the homogeneous solution of (\ref{resc_eqn})) while the second term arises due to rescattering processes (the particular solution 
of (\ref{resc_eqn})).  The bump-like feature in the power spectrum (\ref{Pfit}) will be associated, also, with localized non-Gaussian 
features in the $n$-point correlation functions.

The polyspectra for the model (\ref{Lresc}) are very far from scale invariant, reflecting the fact that rescattering effects are most 
important for modes leaving the horizon near the moment when $\phi=\phi_0$.  To compute the cumulants in this model, we smooth on scales 
$R_b \sim k_b^{-1}$, corresponding to the location of the bump in (\ref{Pfit}).  The calculation is carried out in detail in Appendix C.  
Here we simply quote the key result: 
\begin{equation}
\label{resc_cum}
  \langle \zeta^n({\bf x}) \rangle_{R_b} \sim a_n \left[ (2\pi g) \, \frac{H^2}{2\pi |\dot{\phi}|} \right]^{n}
\end{equation}
where $a_n$ are dimensionless coefficients; more on these shortly.  The variance of fluctuations gets important contributions from both rescattering and also the usual vacuum fluctuations.  We can distinguish two regimes, depending on which effect dominates:
\begin{equation}
 \langle \zeta^2({\bf x}) \rangle_{R_b} \sim \left\{  
  \begin{array}{ll}
    \frac{H^4}{(2\pi)^2\dot{\phi}^2}, & \hspace{5mm}\mbox{vacuum\hspace{2mm}dominated} \\
    a_2 \left(2\pi g\right)^2 \frac{H^4}{(2\pi)^2\dot{\phi}^2} & \hspace{5mm}\mbox{rescattering\hspace{2mm}dominated}
  \end{array}
 \right.
\label{resc_var}
\end{equation}
The dimensionless moments are defined in position space as 
$\mathcal{M}_n = \frac{\langle \zeta^n({\bf x}) \rangle}{\langle \zeta^2({\bf x})\rangle^{n/2}}$.  We find the following result
\begin{equation}
\label{MomentsResc}
   \mathcal{M}_n^{\mathrm{resc}} \sim \left\{  
  \begin{array}{ll}
    a_n\left( 2\pi g \right)^n \, , & \hspace{5mm}\mbox{vacuum\hspace{2mm}dominated} \\
    a_n (a_2)^{-n/2} \, , & \hspace{5mm}\mbox{rescattering\hspace{2mm}dominated}
  \end{array}
 \right.
\end{equation}
We see a structure very much analogous to (\ref{MomentsInvDec}), which was obtained for inverse decay.  However, unlike the previous case there is an extra dependence on model parameters in the prefactor: $a_n = \left(\frac{g\dot{\phi}}{H^2}\right)^{3/2} A_n$, where $A_n$ is a numerical coefficient, independent of model parameters.  Taking this into account, we have an extra parametric dependence in the vacuum dominated regime: $\mathcal{M}_n \propto A_n \; \mathcal{C}\; \left(2\pi g\right)^n$ where $\mathcal{C} \equiv \left(\frac{g\dot{\phi}}{H^2}\right)^{3/2}$.  In the rescattering dominated regime, we instead have $\mathcal{M}_n \propto \left(\mathcal{C}^{-1} \right)^{\frac{n-2}{2}}$.  In the case at hand $\mathcal{C} > 1$ and so the moments are ordered.  The situation is reminiscent of (\ref{salvage}) where the moments fall off at a rate similar to the hierarchical case, however, the small parameter is decoupled from the amplitude of the variance. 

The regime where rescattering effects dominate the variance (\ref{resc_var}) is under computational control.  The bump-like feature in 
(\ref{Pfit}) can dominate over the usual scale-invariant contribution, even while the coupling is perturbatively small ($g^2 < 1$) and 
backreaction effects are negligible \cite{Barnaby:2010sq}.  Note that nothing forbids the inclusion of a bare mass term 
$\mu^2\chi^2$ in the Lagrangian (\ref{Lresc}); even if the $\chi$ field is classically massless at $\phi=\phi_0$ one expects that such a term
will be generated by radiative corrections.  The new parameter $\mu$ has the effect of reducing the efficiency of particle production effects.
For models obtained from super-gravity or string theory it is natural to have $\mu \sim H$.  In this case the polyspectra are essentially
unsuppressed and our analysis is not modified in any important way \cite{Barnaby:2010ke}.

The analysis presented in this section suggests that our qualitative results for the pattern of moments may be rather generic for any 
realization of the feeder mechanism.  It would be interesting to confirm this intuition explicitly.   

\section{Observational Implications of the Structure of Correlation Functions}
\label{sec:observational}

The different scaling of the moments above means that the physical origin of non-Gaussianity may only be evident by looking at observables beyond the bispectrum. The four-point function in the CMB has been studied for nearly a decade \cite{Okamoto:2002ik, Kogo:2006kh} and recently constraints at an interesting level have been obtained. For local type non-Gaussianity four different techniques \cite{Vielva:2009jz, Smidt:2010sv, Fergusson:2010gn, Rossmanith:2011da} have been used to constrain the amplitude to a level where the scaling of the moments might soon be checked. In addition, Fergusson et al \cite{Fergusson:2010gn} apply their technique to also constrain other trispectra, including the equilateral type. If we know which shapes of trispectra and bispectra to compare from particular theoretical models, direct comparison of the amplitudes of those moments is a promising direction, especially given the expected sensitivity of the Planck satellite \cite{Kogo:2006kh}. In this section we investigate a different approach, demonstrating how cross-checks of Gaussianity like Minkowski functionals and cluster number counts (which use integrated moments rather than momentum space $n$-point functions) are sensitive to the structure of the correlation functions.

\subsection{Summary of Scalings}

\begin{figure}[h]
\begin{center}
\includegraphics[width=0.9\textwidth,angle=0]{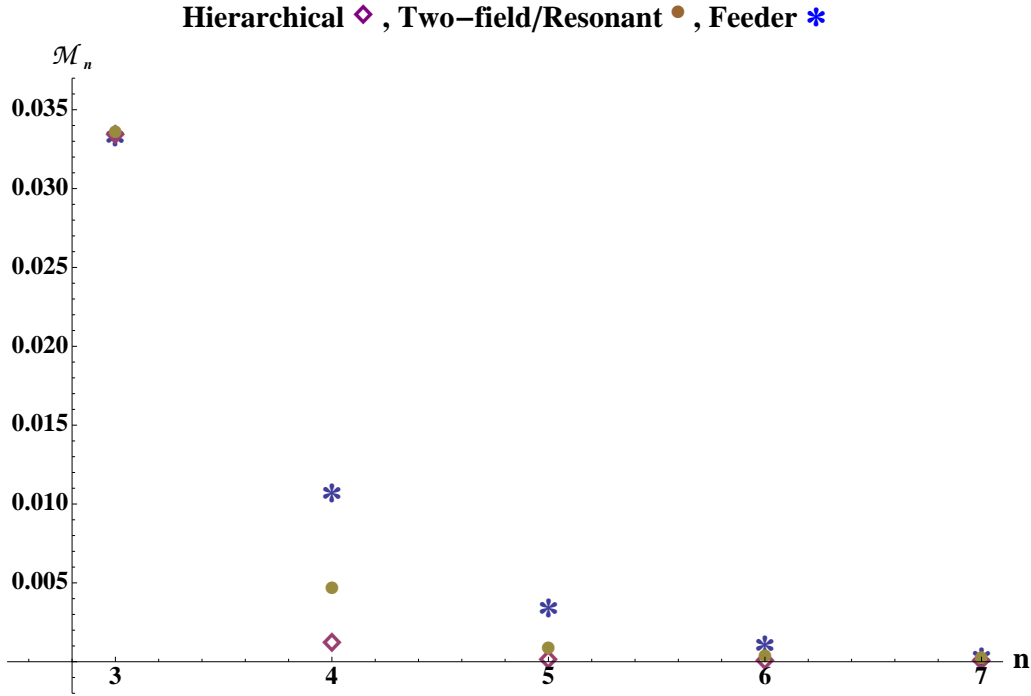} 
\caption{A comparison of the scaling behaviors. All cases have been normalized to give the same value of $\mathcal{M}_3$. The purple diamonds show the standard hierarchical scaling, generated by derivative self-interactions and the local ansatz, for example. For a fixed amplitude of the bispectrum, scenarios of this type are the least non-Gaussian of the scalings discussed here. The brown circles show a hierarchical scaling with an extra parametric dependence in the individual moments, as occurs in resonant non-Gaussianity or the mixed inflaton/curvaton scenario. This last case should also mimic some of the difference that might come from having very different $A_n$ factors. The blue stars show the feeder field scaling given for example from the inverse decay, particle production and ``un-Gaussiton" scenarios. }
\label{fig:DimLessMoments}
\end{center}
\end{figure}

Before considering the smoothed, real-space moments we summarize our findings based on the analysis of momentum space amplitudes of the 
polyspectra. We find two particularly important patterns in the dimensionless moments in terms of the parameters $\mathcal{I}$ labeling the interaction strength and the amplitude of fluctuations $\mathcal{P}_{\zeta}$. One scaling is a characteristic of single-field interactions, called the hierarchical scaling. The other is generated by what we call the feeder mechanism and is characteristic of interactions between the 
inflaton and a second field that is a spectator in terms of the inflationary background. The scalings are captured 
by the ratio of consecutive dimensionless moments, but the moments themselves may have amplitudes shifted by numerical coefficients 
or parametric coefficients that do not scale with order $n$. These are:
\begin{itemize}
 \item {\bf Hierarchical scaling} (local model, derivative self-interactions):
 \[
    \boxed{\mathcal{M}_n \propto A_n\left(\mathcal{I}^22\pi^2\mathcal{P}_{\zeta}\right)^{\frac{n-2}{2}}}
 \]
 \item {\bf Hierarchical with parametric coefficient} (resonance self-interactions):
 \[
    \boxed{\mathcal{M}_n \propto A_n\;\mathcal{C}\;\left(\mathcal{I}^22\pi^2\mathcal{P}_{\zeta}\right)^{\frac{n-2}{2}}}
 \]
 \item {\bf Feeder scaling} (inverse decay):
 \[
   \boxed{\mathcal{M}_n \propto A_n\;\mathcal{I}^{\,n}}
 \]
 \item {\bf Feeder with parametric coefficient} (massive particle production):
 \[
   \boxed{\mathcal{M}_n \propto A_n\;\mathcal{C}\;\mathcal{I}^{\,n}}
 \]
\end{itemize}
We illustrate the first three scalings in Figure \ref{fig:DimLessMoments}, where for visualization purposes parameters have been chosen so that 
the numerical value of $\mathcal{M}_3$ is equal for all scenarios (but unrealistically large) and all the $A_n=1$ for simplicity.

Above we have focused on the simplest scenario where the contribution of the feeder field to the power spectrum is sub-dominant.  In the opposite regime the moments are not generically ordered and a systematic expansion of the PDF around a Gaussian is not possible.  Moreover, the non-ordered regime is ruled out observationally in the simplest models. 

\subsection{Mass Function}

Perhaps the most interesting implication of the different scaling of the moments is in what we would predict for the non-Gaussian mass 
function of gravitationally collapsed objects. Following the Press-Schechter approach \cite{Press:1973iz} (which has previously been 
extended to non-Gaussian initial conditions in \cite{Chiu:1997xb, Robinson:1999se, Matarrese:2000iz, LoVerde:2007ri, LoVerde:2011iz}), to 
predict the number density $dn(M)dM$ of halos with masses in the range $(M,M+dM)$ we need an expression for the non-Gaussian probability 
distribution of (normalized) density fluctuations smoothed on a scale $R$ associated to halos of mass $M$ by 
$R=\left(\frac{3}{4\pi\rho}M\right)^{1/3}$. Labeling the distribution $P(\nu,M)$ with $\nu=\delta/\sigma_R$ -- here $\sigma_R$ is the smoothed variance defined, explicitly below -- the fraction of volume in collapsed objects (halos) is
\be
\label{eq:collapseFraction}
F(M)=\int_{\delta_c/\sigma_R}^\infty\!\!d\nu P(\nu,M)
\ee
with $\delta_c$ the threshold for collapse.  The mass function is
\be
\frac{dn}{dM}(M,z)=-2\frac{\bar{\rho}}{M}\frac{dF}{dM}\;,
\label{eq:dndmdef}
\ee
where $\bar{\rho}=\Omega_m\rho_{crit}$ is the average (co-moving) matter density. The smoothed density field is given by
\be
\delta_R(z)=\int \frac{d^3k}{(2\pi)^3} W_R(k)\delta(\vec{k},z)
\ee
where $W_R(k)$ is the Fourier transform of a window function, commonly taken to be a top-hat in real space giving
\be
W_R(k)=\frac{3\sin(kR)}{k^3R^3}-\frac{3\cos(kR)}{k^2R^2}.
\ee
The relation between the primordial curvature perturbation $\zeta$ and the linear perturbation to the matter density $\delta =\delta \rho/\rho$ today is
\ba
\delta(\vec{k},z)&=&M(k,z)\zeta(\vec{k})\nn
M(k,z)&=&\frac{2}{5}\frac{1}{\Omega_{m}}\frac{1}{H_0^2}D(z)T(k)k^2
\label{Mka}
\ea
where $D(z)$ is the linear growth function, $z$ is the redshift, and $T(k)$ is the transfer function. The smoothed variance and 3-point are, for example,
\ba
\label{eq:smoothsig}
\sigma^2(R)=\langle\delta^2_R\rangle&=&\int\frac{d^3k}{(2\pi)^3}\int\frac{d^3{ k^{\prime}}}{(2\pi)^3} W_R(k)W_R(k')M(k,z)M(k',z)\langle\zeta(\vec{k})\zeta(\vec{k}')\rangle\nn
&=&\int_0^\infty \frac{dk}{k} W_R(k)^2M(k,z)^2 \mathcal{P}_\zeta(k)\\\nonumber
\langle\delta^3_R\rangle&=&\int\frac{d^3k_1}{(2\pi)^3}\int\frac{d^3k_2}{(2\pi)^3}\int\frac{d^3k_3}{(2\pi)^3}W_1W_2W_3M_1M_2M_3\langle{\zeta(\vec{k}_1)\zeta(\vec{k}_2)\zeta(\vec{k}_3)}\rangle
\ea

To find an approximate expression for the non-Gaussian probability density, we will use the Edgeworth expansion. Although one can think of various ways to express a slightly non-Gaussian distribution as a series (the Gram-Charlier and Gauss-Hermite expansions are two others) the Edgeworth expansion is neatly organized by the moments of the distribution and is a true asymptotic expansion, allowing one to calculate the error when the series is truncated. The expansion was originally developed to describe distributions that approach the Gaussian (according to the central limit theorem) as the number of degrees of freedom, $n$, increases. In this case the expansion is organized in powers of $1/\sqrt{n}$. Although only the first few terms are typically quoted in the physics literature, Petrov \cite{Petrov:1972, Petrov:1987} (references are the English translations) developed an expression for the complete series and proved the expansion is asymptotic for cases where $\sigma\sim1/\sqrt{n}$. 

The common hierarchical scaling often prompts a definition of the ``reduced cumulants" 
 \be
 S_{n,R}\equiv\frac{\langle\delta_R^n\rangle_c}{\langle\delta_R^2\rangle_c^{n-1}}
 \ee
 which are related to the dimensionless moments by
 \be
 \mathcal{M}^h_{n,R}=S_{n,R}\sigma_R^{n-2}\;.
 \ee
Note that the dimensionless moments are always redshift independent. Then Petrov's expression, written in terms of (a generic) $\sigma$ rather than $\sqrt{n}$ and in terms of the smoothed, reduced cumulants, is
\be
\label{eq:Petrov}
P(\nu){\rm d}\nu=\frac{d\nu}{\sqrt{2\pi}}e^{-\nu^2/2}\left\{1+\sum_{s=1}^{\infty}\sigma_R^s\sum_{\{k_m\}}{\rm H}_{s+2r}(\nu)\prod_{m=1}^2\frac{1}{k_m!}\left(\frac{S_{m+2,R}}{(m+2)!}\right)^{k_m}\right\}\;.
\ee
Here $H_n(\nu)$ are Hermite polynomials defined by $H_n(\nu)=(-1)^ne^{\nu^2/2}\frac{{\rm d}^n}{{\rm d}\nu^n}e^{-\nu^2/2}$ and $r=k_1+k_2+\dots+k_n$ where the set $\{k_m\}$ is built of all non-negative integer solutions of
\be
\label{eq:Diophantine}
k_1+2k_2+\dots+nk_n=s\;.
\ee

We can re-write this expression in a form that will allow us to organize the expansion appropriately for non-hierarchical scalings by using the dimensionless smoothed moments $\mathcal{M}_{n,R}$. Making use of Eq.(\ref{eq:Diophantine}), we find
\ba
\label{eq:PetrovM}
P(\nu){\rm d}\nu&=&\frac{d\nu}{\sqrt{2\pi}}e^{-\nu^2/2}\left\{1+\sum_{s=1}^{\infty}\sum_{\{k_m\}}{\rm H}_{s+2r}(\nu)\prod_{m=1}^2\frac{1}{k_m!}\left(\frac{\mathcal{M}_{m+2,R}}{(m+2)!}\right)^{k_m}\right\}\\\nonumber
&\equiv&\frac{d\nu}{\sqrt{2\pi}}e^{-\nu^2/2}(1+p_1(\nu,R)+p_2(\nu,R)+\dots)\;
\ea
where $p_i(\nu,R)$ organize the expansion in terms of the small parameter which controls the strength of interactions.  As we motivated for several examples in earlier sections, it is the smallness of the $\mathcal{M}_n$ that controls the departure from Gaussianity, and so to keep good control of the errors in the expansion we should group terms of the same order according to the scaling of the moments.

For the hierarchical scaling, where it is natural to group the $\mathcal{M}^h_{4,R}$ term with the $(\mathcal{M}^h_{3,R})^2$ term, etc, this leads to the standard grouping in powers of $\sigma$ ($\sim1/\sqrt{n}$ for $n$ degrees of freedom) \cite{LoVerde:2007ri}. Then the first few terms look like
\ba
\label{EdgeworthHier}
p_1^{(h)}(\nu,R)&=&\frac{\mathcal{M}^h_{3,R}}{3!}H_3(\nu)\\\nonumber
p_2^{(h)}(\nu,R)&=&\frac{\mathcal{M}^h_{4,R}}{4!}H_4(\nu)+\frac{1}{2}\left(\frac{\mathcal{M}^h_{3,R}}{3!}\right)^2H_6(\nu)\\\nonumber
p_3^{(h)}(\nu,R)&=&\frac{\mathcal{M}^h_{5,R}}{5!}H_5(\nu)+\frac{\mathcal{M}^h_{3,R}\mathcal{M}^h_{4,R}}{3!4!}H_7(\nu)+\frac{1}{3!}\left(\frac{\mathcal{M}^h_{3,R}}{3!}\right)^3H_9(\nu)
\ea

On the other hand, for the feeder field type scaling we expect the $(\mathcal{M}^f_{3,R})^2$ term to be roughly of the same order as the $\mathcal{M}^f_{6,R}$ term, etc. In that case, we have
\ba
\label{EdgeworthAxion}
p_1^{(f)}(\nu,R)&=&\frac{\mathcal{M}^f_{3,R}}{3!}H_3(\nu)\\\nonumber
p_2^{(f)}(\nu,R)&=&\frac{\mathcal{M}^f_{4,R}}{4!}H_4(\nu)\\\nonumber
p_3^{(f)}(\nu,R)&=&\frac{\mathcal{M}^f_{5,R}}{5!}H_5(\nu)\\\nonumber
p_4^{(f)}(\nu,R)&=&\left(\frac{\mathcal{M}^f_{6,R}}{6!}+\frac{1}{2}\left(\frac{\mathcal{M}^f_{3,R}}{3!}\right)^2\right)H_6(\nu)\\\nonumber
p_4^{(f)}(\nu,R)&=&\left(\frac{\mathcal{M}^f_{3,R}\mathcal{M}^f_{4,R}}{3!4!}+\frac{\mathcal{M}^f_{7,R}}{7!}\right)H_7(\nu)
\ea
We should mention that the integrations smoothing over the momentum space $n-1$ spectra (as in Eq.(\ref{eq:smoothsig})) appear to change the scaling of the moments slightly (eg, see Figure 1 in \cite{LoVerde:2011iz} where $\mathcal{M}_{4,R}\approx 2.5\mathcal{M}_{3,R}^2$). One can probably account for this systematically, and in any case that contribution to the scaling will be the same for all scenarios, regardless of how the coefficients of the momentum space spectra scale. Since both the hierarchical and feeder field scenarios would shift in the same way, this should not qualitatively affect our comparison of the two cases.

We note that expansions following from Eq.(\ref{eq:PetrovM}) need only resemble that of the actual non-Gaussian distribution near the peak, with a smaller range of validity onto the tail the more the actual distribution deviates from a Gaussian (see \cite{Blinnikov:1997jq} for some illustrative examples). However, the point where the errors are large {\it can} be computed by looking at the size of the terms in the series. From this we also see that the inverse-decay dominated regime, where all the $\mathcal{M}_n$ are the same size, is not well-captured by any truncation of this expansion except very near the peak ($\nu\ll1$).

Using these probability distributions in Eq.(\ref{eq:dndmdef}), we can expand the collapse fraction
\be
F(M)=F_0(M)+F_1(M)+F_2(M)+\dots
\ee
where 
\be
F_0(M)=\frac{1}{2}{\rm Erfc}\left(\frac{\nu_c}{\sqrt{2}}\right)
\ee
and the non-Gaussian terms $F_1(M)$ etc contain dimensionless moments grouped at each order according to their scaling. To write the mass function we will need derivatives of those terms with respect to mass (or smoothing scale). The first term is the usual Gaussian, so regardless of non-Gaussian type we have
\be
F_0^{\prime}\equiv\frac{dF_0}{dM}=\frac{e^{-\frac{\nu_c^2}{2}}}{\sqrt{2\pi}}\frac{d\sigma}{dM}\;\frac{\nu_c}{\sigma}
\ee
Then the ratio of the non-Gaussian Edgeworth mass function to the Gaussian has the same structural form for either scaling:
\be
\label{EdgeworthMassfcn}
\frac{n_{NG}}{n_G}|_{{Edgeworth}}\approx1+\frac{F^{(h,f)\prime}_1(M)}{F^{\prime}_0(M)}+\frac{F^{(h,f)\prime}_2(M)}{F^{\prime}_0(M)}+\dots
\ee

In the next section we use a specific example to show how the difference in mass function might help us distinguish the two scenarios observationally. Although we have not explicitly written all the terms above, we will show the results up to third order for for the feeder field case to show how the series is behaving.

For reference, the first few Hermite polynomials are:
\ba
\label{eq:Hermites}
H_0(\nu)&=&1\\\nonumber
H_1(\nu)&=&\nu\\\nonumber
H_2(\nu)&=&\nu^2-1\\\nonumber
H_3(\nu)&=&\nu^3-3\nu\\\nonumber
H_4(\nu)&=&\nu^4-6\nu^2+3\\\nonumber
H_5(\nu)&=&\nu^5-10\nu^3+15\nu\\\nonumber
H_6(\nu)&=&\nu^6-15\nu^4+45\nu^2-15\\\nonumber
H_7(\nu)&=&\nu^7-21\nu^5+105\nu^3-105\nu\\\nonumber
H_8(\nu)&=&\nu^8-28\nu^6+210\nu^4-420\nu^2+105\\\nonumber
H_9(\nu)&=&\nu^9-36\nu^7+378\nu^5-1260\nu^3+945\nu.
\ea

\subsection{Comparing Hierarchical Scaling to the Feeder Mechanism}
\label{subsec:comparing}

As a reference point, let's assume that a bispectrum of the equilateral shape has been detected. That is, we have measured the amplitude, $\mathcal{B}_{eq}$, using the equilateral template \cite{Creminelli:2006rz} (for simplicity we assume $n_s=1$)
\ba
\label{eqShape}
\Expect{\zeta(\vec{k}_1)\zeta(\vec{k}_2)\zeta(\vec{k}_3)}&=&(2\pi)^3\delta^3_D(\vec{k}_1+\vec{k}_2+\vec{k}_3)\;\mathcal{B}_{eq}\\\nonumber
&&\times\left[-\frac{1}{(k_1k_2)^3}+\textrm{2 perm.}-\frac{2}{(k_1k_2k_3)^2}+\frac{1}{k_1k_2^2k_3^3}+\textrm{5 perm.}\right]\;.
\ea
Suppose we find $\mathcal{B}_{eq}=3.3\times10^{-15}$, corresponding to an effective nonlinearity parameter $f_{NL}^{eq}=\mathcal{B}_{eq}/(6\mathcal{P}_{\zeta}^2)\approx100$. If we attribute this to a model with hierarchical scaling, then we expect $\mathcal{M}^{h}_4\sim \mathcal{O}(\mathcal{M}^{h}_3)^2$.  On the other hand, if we attribute it to a feeder field (eg - inverse decay) then we expect $\mathcal{M}^{f}_4\sim \mathcal{O}(\mathcal{M}^f_3)^{4/3}$ (ignoring for the moment the slight discrepancy in the exact shapes from the two scenarios). Clearly, to distinguish these scenarios we need more information. This could come from a measurement of the trispectrum,\footnote{The precise shape of the bispectrum and the tensor-to-scalar ratio may also be a valuable discriminators.} but here we will explore a consistency check from the distribution of high mass/high redshift clusters. 

Since we have measured the three-point in our scenario the first non-Gaussian term in the mass function is determined. To distinguish the scenarios we will compare what we would predict for the non-Gaussian mass functions at next order. That is, we will compare the second order mass function with hierarchical scaling to the second order mass function with feeder field type scaling.

We need the non-Gaussian terms for each type of scaling. For the hierarchical case
\ba
F_1^{(h)}(M)&=&\frac{e^{-\frac{\nu_c^2}{2}}}{\sqrt{2\pi}}\frac{\mathcal{M}^{(h)}_{R,3}}{3!}H_2(\nu_c)\\\nonumber
F_2^{(h)}(M)&=&\frac{e^{-\frac{\nu_c^2}{2}}}{\sqrt{2\pi}}\left[\frac{\mathcal{M}^{(h)}_{R,4}}{4!}H_3(\nu_c)+\frac{1}{2}\left(\frac{\mathcal{M}^{(h)}_{R,3}}{3!}\right)^2H_5(\nu_c)\right]
\ea
and for the feeder field
\ba
F_1^{(f)}(M)&=&\frac{e^{-\frac{\nu_c^2}{2}}}{\sqrt{2\pi}}\frac{\mathcal{M}^{(f)}_{R,3}}{3!}H_2(\nu_c)\\\nonumber
F_2^{(f)}(M)&=&\frac{e^{-\frac{\nu_c^2}{2}}}{\sqrt{2\pi}}\frac{\mathcal{M}^{(f)}_{R,4}}{4!}H_3(\nu_c)\;.
\ea
For a scale invariant equilateral shape, the dimensionless moments $\mathcal{M}_{3,R}$, $\mathcal{M}_{4,R}$ are only very weakly dependent on the smoothing scale. Then the derivatives of the non-Gaussian terms simplify:
\be
F_i^{\prime}\approx-\frac{\delta_c}{\sigma^2}\frac{dF_i}{d\nu}\frac{d\sigma}{dM}
\ee
and so the hierarchical and feeder field terms are
\ba
F_1^{(h)\prime}&\approx&\frac{\mathcal{M}^{(h)}_{R,3}}{3!}\frac{e^{-\frac{\nu_c^2}{2}}}{\sqrt{2\pi}}\frac{d\sigma}{dM}\frac{\nu_c}{\sigma}H_3(\nu_c)=F_0^{\prime}\;\frac{\mathcal{M}^{(h)}_{R,3}}{3!}H_3(\nu_c)\\\nonumber
F_2^{(h)\prime}&\approx&F_0^{\prime}\;\left[\frac{\mathcal{M}^{(h)}_{R,4}}{4!}H_4(\nu_c)+\frac{1}{2}\left(\frac{\mathcal{M}^{(h)}_{R,3}}{3!}\right)^2H_6(\nu_c)\right]\\\nonumber
F_1^{(f)\prime}&\approx&F_0^{\prime}\;\frac{\mathcal{M}^{(f)}_{R,3}}{3!}H_3(\nu_c)\\\nonumber
F_2^{(f)\prime}&\approx&F_0^{\prime}\;\frac{\mathcal{M}^{(f)}_{R,4}}{4!}H_4(\nu_c)
\ea

\begin{figure}[h]
\begin{center}
$\begin{array}{cc}
\includegraphics[width=0.5\textwidth,angle=0]{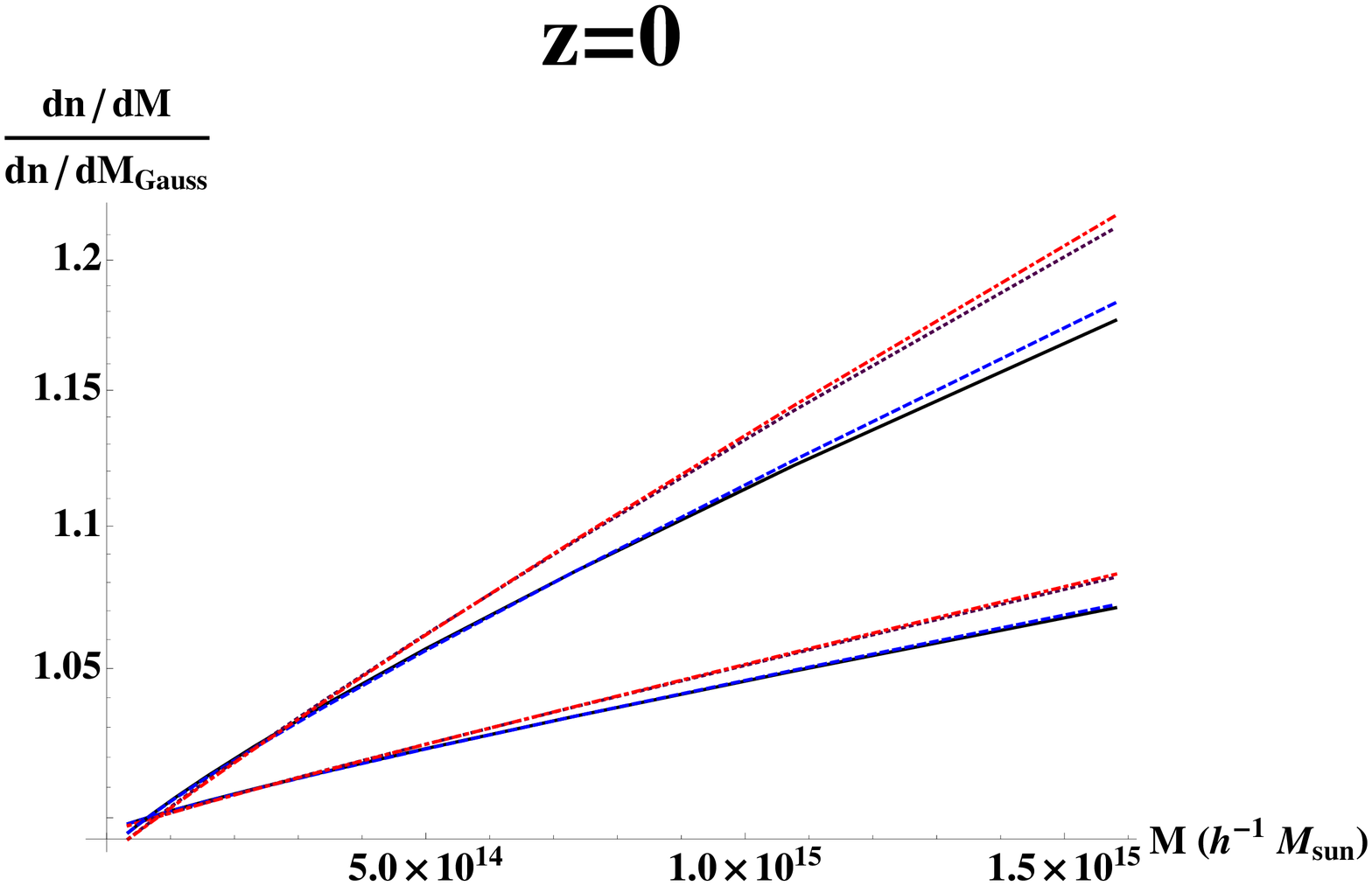} &
\includegraphics[width=0.5\textwidth,angle=0]{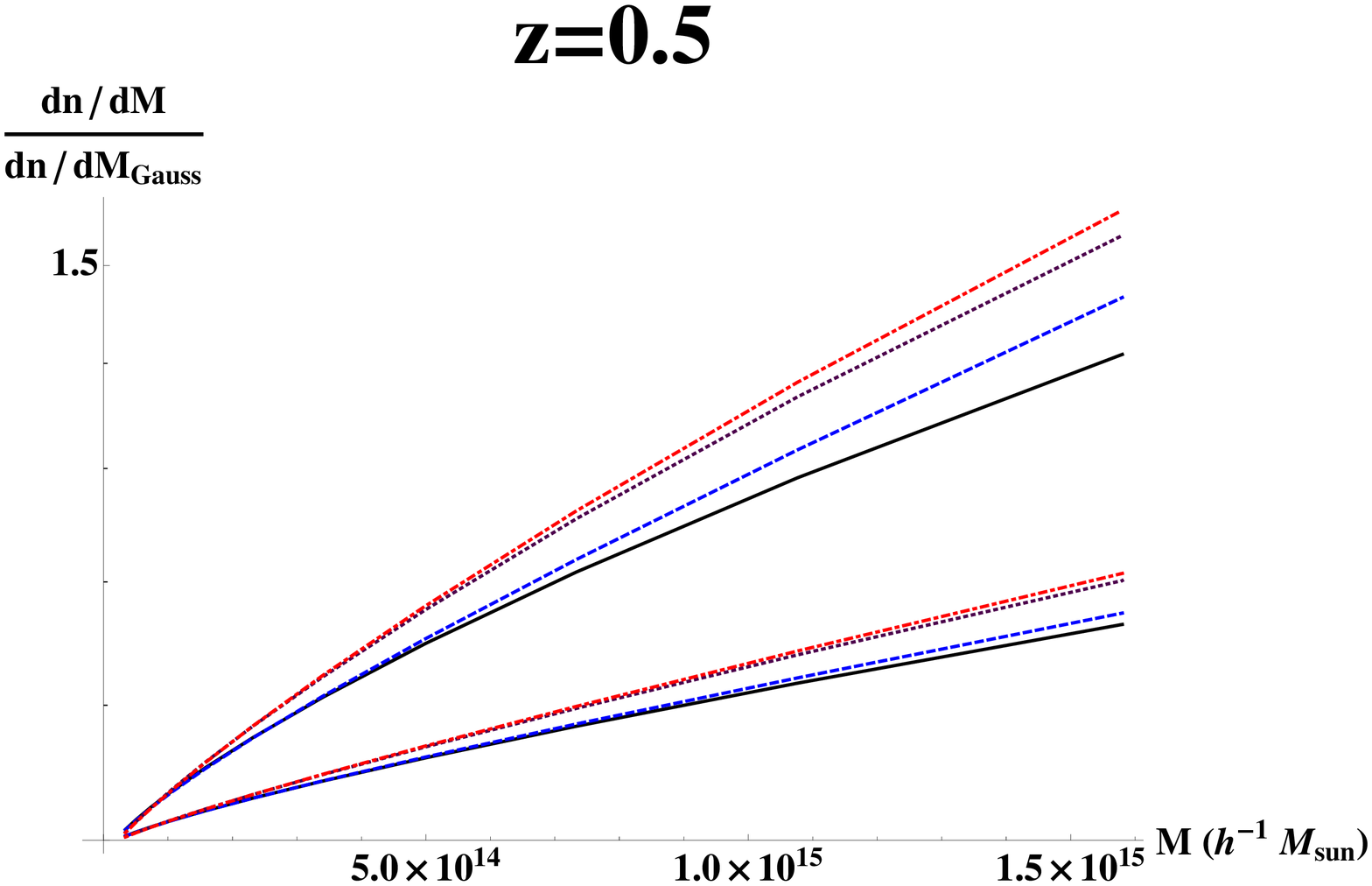} 
\end{array}$
\caption{The ratio of the non-Gaussian to Gaussian mass function for redshifts $z=0$ and $z=0.5$. Each panel shows a set of lower curves that assume equilateral non-Gaussianity corresponding to $f_{NL}=100$ and upper curves corresponding to $f_{NL}=250$. In each set of curves, the black solid lines shows the result for the pdf truncated at $\mathcal{M}_{3,R}$ while the blue dashed and purple dotted lines show the second order result for the hierarchical and feeder field scalings respectively. Notice that the range of masses where the fourth moment is a significant correction is model dependent. Finally, the red dot-dashed lines shows the next order ($\mathcal{M}_{5,R}$) correction for the feeder field scenario.}
\label{fig:compareMF}
\end{center}
\end{figure}

Putting these expressions into Eq.(\ref{EdgeworthMassfcn}) gives the Edgeworth prediction for the non-Gaussian correction factor to the mass function. Figure \ref{fig:compareMF} shows the ratio of the non-Gaussian mass functions to the Gaussian for the two scenarios with the approximation where the terms proportional to derivatives of the $\mathcal{M}_{n,R}$ are dropped. 

From Fig.~\ref{fig:compareMF}, there are two relevant differences between the scenarios. First, in the feeder field case the correction from moments above the third one is relevant at significantly lower mass. Second, the hierarchical scaling predicts fewer high mass/high redshift objects than the non-hierarchical scaling, although at a level that might require futuristic data sets to distinguish. 

A template for an equilateral type trispectrum has been constrained from the CMB by Fergusson et al \cite{Fergusson:2010gn}. The template, chosen to match (up to a numerical factor) one of several explicit single-field shapes calculated by Chen et al \cite{Chen:2009bc}, is
\ba
\langle\zeta(\vec{k}_1)\zeta(\vec{k}_2)\zeta(\vec{k}_3)\zeta(\vec{k}_4)\rangle&=&(2\pi)^3\delta^3_D(\vec{k}_1+\vec{k}_2+\vec{k}_3+\vec{k}_4)\;P_{4,\zeta}(k_1,k_2,k_3,k_4)\\\nonumber
\langle\zeta(\vec{k}_1)\zeta(\vec{k}_2)\zeta(\vec{k}_3)\zeta(\vec{k}_4)\rangle&=&(2\pi)^3\delta^3_D(\vec{k}_1+\vec{k}_2+\vec{k}_3+\vec{k}_4)\;8t_{NL}(2\pi^2\mathcal{P}_{\zeta})^3\:\frac{1}{K^5}\prod_{i=1}^4\frac{1}{k_i}\\\nonumber
K&=&(k_1+k_2+k_3+k_4)/4
\ea
Ref.~\cite{Fergusson:2010gn} quotes the constraint:
\be
t_{NL}=(-3.11\pm7.5)\times 10^6\;.
\ee
From this constraint one can work outs how a particular small sound speed model like DBI is constrained compared to the inverse decay scenario.

\subsection{Log-Edgeworth Expansion}

Although the Edgeworth expansion of the probability distribution has several useful features, it leads to a mass function that is often not well-behaved. Recently LoVerde and Smith \cite{LoVerde:2011iz} have proposed a slightly different expansion technique which leads to what they call the `Log-Edgeworth' mass function\footnote{A different technique for improving the behavior of the series was proposed in \cite{Paranjape:2011uk}, and one might try to similarly extend it to non-hierarchical scalings.}. They compare this mass function and the standard Edgeworth expression to simulations of local-type non-Gaussianity. One of their results is particularly interesting for our purposes: the Log-Edgeworth mass function is in better agreement with simulations that have a local type trispectrum with amplitude greater than (the natural size) $\mathcal{O}(f_{NL}^2)$. Since our feeder field scenario similarly has a boosted four-point amplitude, we compare here the expectation for the mass function using the Log-Edgeworth expansion. Of course, any theoretical mass function must be checked against and calibrated with simulations. Here we anticipate the possibility that the feeder type non-Gaussianity may be better predicted by the Log-Edgeworth expansion in analogy with results found for local-type scenarios that are designed to move away from the hierarchical scaling. 

The Log-Edgeworth mass function comes from truncating the expansion for ${\rm Log}\:F$, the logarithm of the collapse fraction, instead of $F$ \cite{LoVerde:2011iz}. The expression is
\ba
\frac{n_{NG}}{n_G}|_{{\rm log-Edgeworth}}&\approx&{\rm Exp}\left[\frac{F_1(M)}{F_0(M)}+\frac{F_2(M)}{F_0(M)}-\frac{1}{2}\left(\frac{F_1(M)}{F_0(M)}\right)^2\right]\\\nonumber
&&\times\left\{1+\frac{F^{\prime}_1(M)+F^{\prime}_2(M)}{F^{\prime}_0(M)}-\frac{F_1(M)F^{\prime}_1(M)}{F_0(M)F^{\prime}_0(M)}\right.\\\nonumber
&&\left.-\frac{F_1(M)+F_2(M)}{F_0(M)}+\left(\frac{F_1(M)}{F_0(M)}\right)^2\right\}\;.
\ea
This formula works for both the hiearchical or feeder scenarios, provided the correct expression for the $F_i(M)$ are used; see subsection \ref{subsec:comparing}.

\begin{figure}[h]
\begin{center}
$\begin{array}{cc}
\includegraphics[width=0.5\textwidth,angle=0]{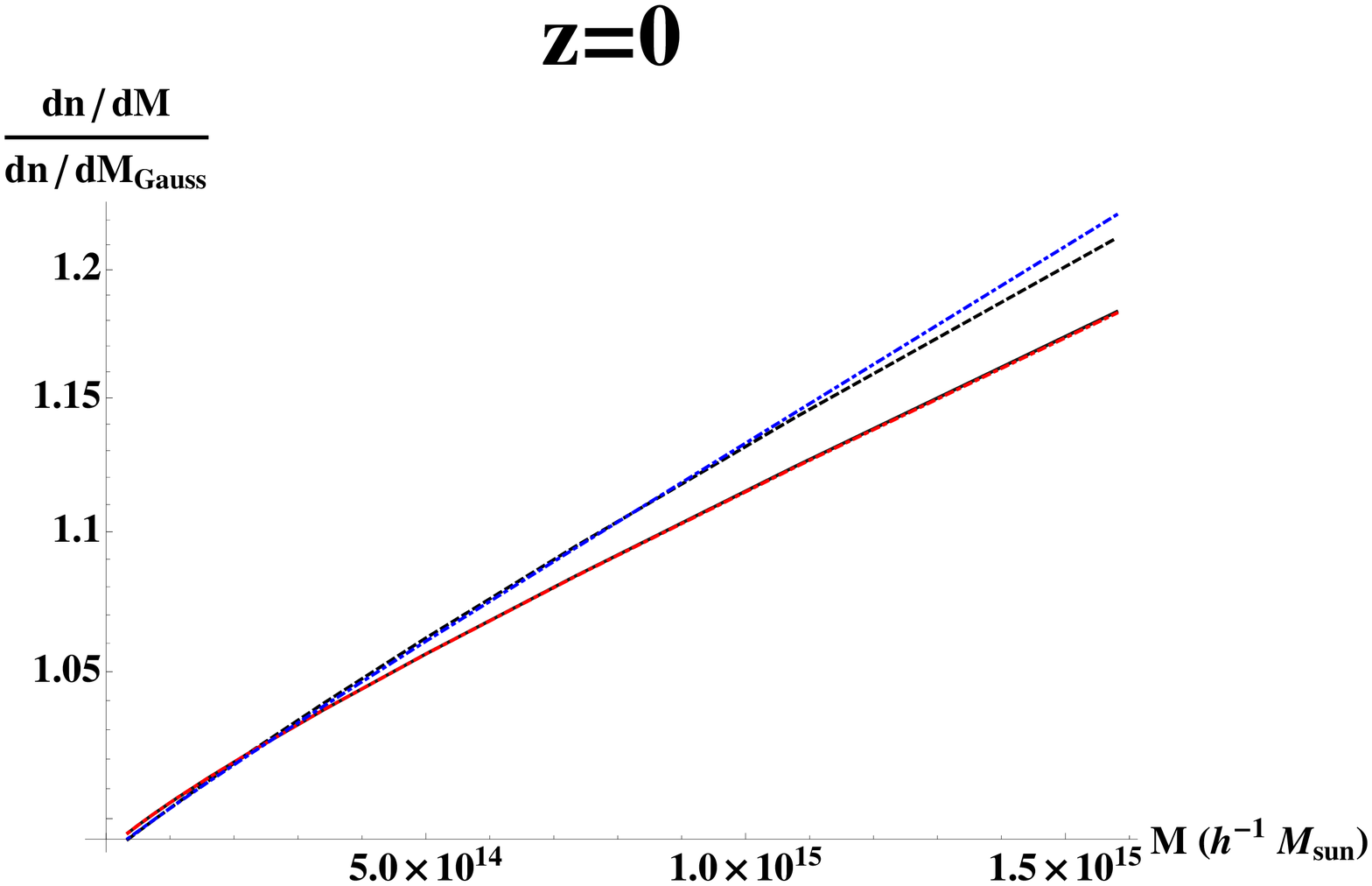} &
\includegraphics[width=0.5\textwidth,angle=0]{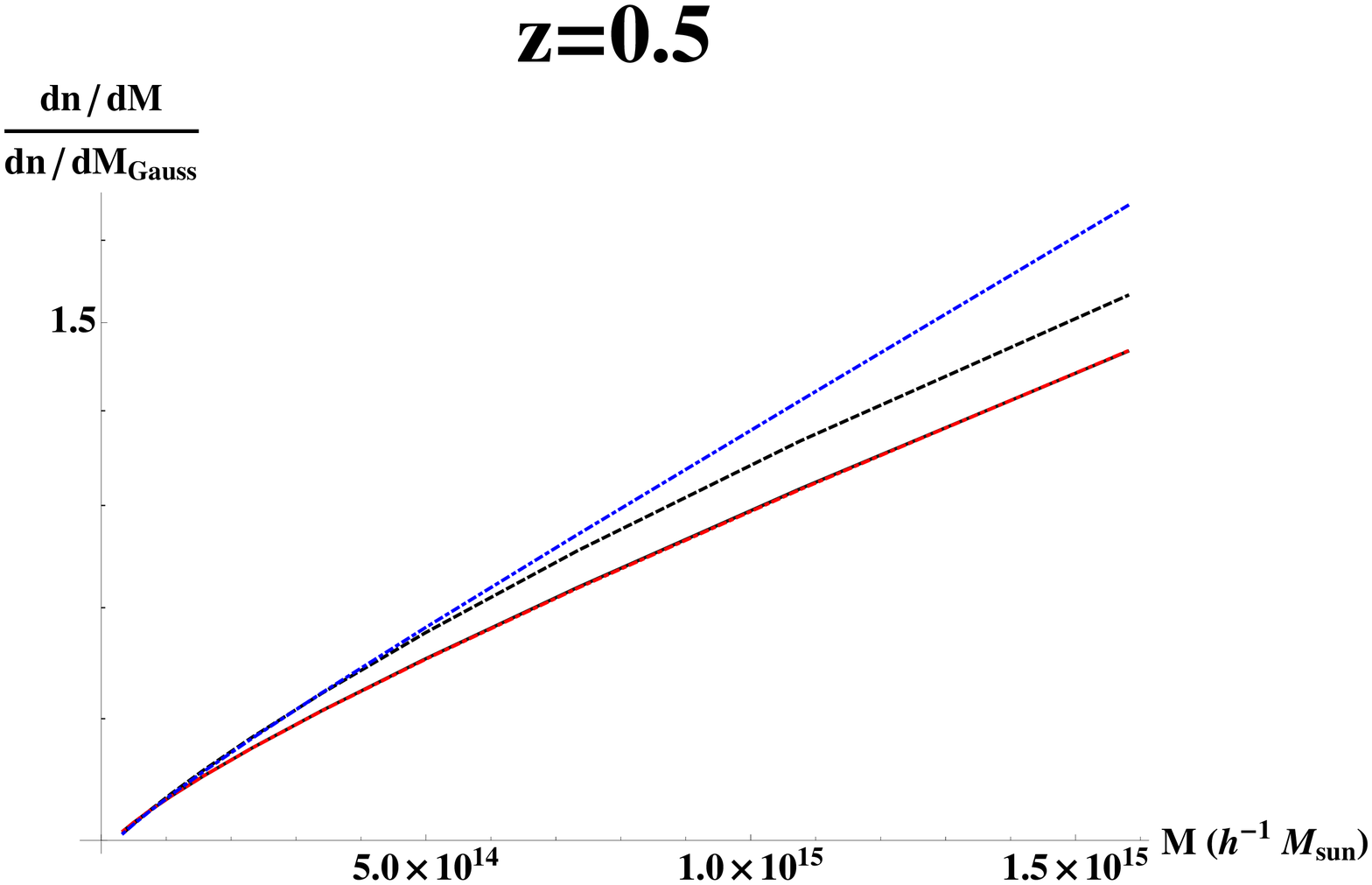} 
\end{array}$
\caption{A comparison of the predictions from the second order Edgeworth and Log-Edgeworth mass functions. The ratio of the non-Gaussian to Gaussian mass functions of each type is plotted versus mass for redshifts $z=0$ and $z=0.5$. Each panel shows the hierarchical (lower curves) and feeder field (upper curves) scaling assuming an equilateral type bispectrum with amplitude corresponding to $f_{NL}^{equil}=250$. The solid black lines show the second order Edgeworth result for hierarchical scaling. The dashed red lines are the Log-Edgeworth expansions for the hierarchical case and are indistinguishable from the Edgeworth for the parameters plotted here. The dashed black lines show the Edgeworth expansion for the feeder field scaling, and the dot-dashed blue lines show the Log-Edgeworth. Interestingly, the second-order Log-Edgeworth for the feeder field corresponds very nearly with the fourth order Edgeworth expansion (see previous figure). }
\label{fig:LogEdgeworth}
\end{center}
\end{figure}

Figure \ref{fig:LogEdgeworth} compares the Edgeworth and Log-Edgeworth mass functions for both scalings. The difference between the two expressions is much more significant for the feeder field type scaling. The Log-Edgeworth result at second order nearly corresponds with the fourth order Edgeworth result, and indicates a larger deviation between the two scalings than is evident at first order.

\subsection{Minkowski Functionals}

Minkowski functionals are morphological indicators of the statistical properties of a generic random field, $f$. They have been used to cross-check constraints of non-Gaussianity from the CMB and for weakly non-Gaussian fields can be approximated by a series in the moments of the distribution much like the Edgeworth expansion. General expressions for the non-Gaussian Minkowski functionals have been worked out in detail by Matsubara \cite{Matsubara:2003yt}. CMB analysis has so far been done for Minkowski Functionals with the first order correction \cite{Hikage:2008gy, Natoli:2009wk}, but Matsubara has recently presented the analytic expressions needed for the second order terms for scenarios with hierarchical scaling \cite{Matsubara:2010te} (see also \cite{Munshi:2010df}).

If $f$ is a two-dimensional field, three Minkowski functionals for the normalized field $\nu=f/\sigma$ measure the fractional area of the regions above threshold ($V_0(\nu_t)$), the length of the boundaries of regions above threshold ($V_1(\nu_t)$), and the number of contours surrounding regions above threshold minus the number of contours surrounding regions below threshold ($V_2(\nu_t)$). These quantities rely on gradients of the field, and so the expansion about the Gaussian will contain several statistics at each order $n$. For models with hierarchical scaling, these have been labeled as 2 variances, 3 reduced skewness parameters, 4 reduced kurtosis parameters, etc. :
\ba
n&=&2:\;\;\;\;\sigma^2=\langle f^2\rangle\;,\;\;\;\;\sigma_1^2=-\langle f\nabla^2f\rangle\\\nonumber
n&=&3:\;\;\;\;S=\frac{\langle f^3\rangle}{\sigma^4}\;,\;\;\;\;S_I=\frac{\langle f^2\nabla^2f\rangle}{\sigma^2\sigma_1^2}\;,\;\;\;\;S_{II}=\frac{2\langle|\nabla f|^2\nabla^2f\rangle}{\sigma_1^4}\\\nonumber
n&=&4:\;\;\;\; K=\frac{\langle f^4\rangle_c}{\sigma^6}\;,\;\;\;\;K_I=\frac{\langle f^3\nabla^2f\rangle_c}{\sigma^4\sigma_1^2}\;,\;\;\;\;K_{II}=\frac{2\langle f|\nabla f|^2\nabla^2f\rangle_c+\langle|\nabla f|^4\rangle_c}{\sigma^2\sigma_1^4}\;,\\\nonumber
&&\;\;\;\;\;\;\;\; \; K_{III}=\frac{\langle|\nabla f|^4\rangle}{2\sigma^2\sigma_1^4}
\ea
To reduce the notational complexity, we only give the explicit expressions relevant for the two-dimensional field of the CMB. In that case, there are three Minkowski functionals, expressed as a function of threshold $\nu_t$. In a form appropriate for the hierarchical scaling, the second order expressions are:
\ba
V_0(\nu_t)|_{{\rm hier}}&=&\frac{1}{\sqrt{2\pi}}e^{-\nu_t^2/2}\left\{e^{\nu_t^2/2}\sqrt{\frac{2}{\pi}}{\rm erfc}\left(\frac{\nu_t}{\sqrt{2}}\right)+\sigma\frac{S}{6}H_2(\nu_t)+\right.\\\nonumber
&&\left.+\sigma^2\left[\frac{S^2}{72}H_5(\nu_t)+\frac{K}{24}H_3(\nu_t)\right]+\dots\right\}\\\nonumber
V_1(\nu_t)|_{{\rm hier}}&=&\frac{1}{8}\frac{\sigma_1}{\sqrt{2}\sigma}e^{-\nu_t^2/2}\left\{1+\sigma\left[\frac{S}{6}H_3(\nu)-\frac{S_{I}}{4}H_1(\nu_t)\right]\right.\\\nonumber
&&\left.+\sigma^2\left[\frac{S^2}{72}H_6(\nu_t)+\frac{K-SS_I}{24}H_4(\nu_t)-\frac{1}{12}\left(K_I+\frac{3}{8}S_I^2\right)H_2(\nu_t)-\frac{K_{III}}{8}\right]+\dots\right\}\\\nonumber
V_2(\nu_t)|_{{\rm hier}}&=&\frac{1}{(2\pi)^{3/2}}\frac{\sigma_1^2}{2\sigma^2}e^{-\nu_t^2/2}\left\{H_1(\nu_t)+\sigma\left[\frac{S}{6}H_4(\nu_t)-\frac{S_I}{2}H_2(\nu_t)-\frac{S_{II}}{2}\right]\right.\\\nonumber
&&\left.+\sigma^2\left[\frac{S^2}{72}H_7(\nu_t)+\frac{K-2S\cdot S_I}{24}H_5(\nu_t)-\frac{1}{6}\left(K_I+\frac{1}{2}S\cdot S_{II}\right)H_3(\nu_t)\right.\right.\\\nonumber
&&\left.\left.-\frac{1}{2}\left(K_{II}+\frac{1}{2}S_I\cdot S_{II}\right)H_1(\nu_t)\right]+\dots\right\}
\ea

As we did for the mass function, we can immediately rewrite the expressions for the Minkowski functionals in a form appropriate for non-hierarchical scaling. First, we will replace the reduced cumulants suitable for the hierarchical scaling by the dimensionless moments:
\ba
n&=&3:\;\;\;\;\mathcal{M}_{3}=\frac{\langle f^3\rangle}{\sigma^3}\;,\;\;\;\;\mathcal{M}_{3}^{(I)}=\frac{\langle f^2\nabla^2f\rangle}{\sigma\sigma_1^2}\;,\;\;\;\;\mathcal{M}_{3}^{(II)}=\frac{2\langle|\nabla f|^2\nabla^2f\rangle}{\sigma_1^3}\left(\frac{\sigma}{\sigma_1}\right)\\\nonumber
n&=&4:\;\;\;\; \mathcal{M}_{4}=\frac{\langle f^4\rangle_c}{\sigma^4}\;,\;\;\;\; \mathcal{M}_{4}^{(I)}=\frac{\langle f^3\nabla^2f\rangle_c}{\sigma^2\sigma_1^2}\;,\;\;\;\; \mathcal{M}_{4}^{(II)}=\frac{2\langle f|\nabla f|^2\nabla^2f\rangle_c+\langle|\nabla f|^4\rangle_c}{\sigma_1^4}\;,\\\nonumber
&&\;\;\;\;\;\;\;\; \;  \mathcal{M}_{4}{(III)}=\frac{\langle|\nabla f|^4\rangle}{2\sigma_1^4}
\ea

Then, the second order expressions for the Minkowski functionals, appropriate for physics with feeder field like scaling are
\ba
\label{eq:MFfeeder}
V_0(\nu_t)|_{{\rm feeder}}&=&\frac{1}{\sqrt{2\pi}}e^{-\nu_t^2/2}\left\{e^{\nu_t^2/2}\sqrt{\frac{2}{\pi}}{\rm erfc}\left(\frac{\nu_t}{\sqrt{2}}\right)+\frac{\mathcal{M}_{3}}{6}H_2(\nu_t)+\right.\\\nonumber
&&\left.+\frac{\mathcal{M}_{4}}{24}H_3(\nu_t)+\dots\right\}\\\nonumber
V_1(\nu_t)|_{{\rm feeder}}&=&\frac{1}{8}\frac{\sigma_1}{\sqrt{2}\sigma}e^{-\nu_t^2/2}\left\{1+\left[\frac{\mathcal{M}_{3}}{6}H_3(\nu_t)-\frac{\mathcal{M}_{3}^{(I)}}{4}H_1(\nu_t)\right]\right.\\\nonumber
&&\left.+\frac{\mathcal{M}_{3}^2}{72}H_6(\nu_t)+\frac{\mathcal{M}_{4}}{24}H_4(\nu_t)-\frac{\mathcal{M}_{4}^{(I)}}{12}H_2(\nu_t)-\frac{\mathcal{M}_{4}^{(III)}}{8}+\dots\right\}\\\nonumber
V_2(\nu_t)|_{{\rm feeder}}&=&\frac{1}{(2\pi)^{3/2}}\frac{\sigma_1^2}{2\sigma^2}e^{-\nu_t^2/2}\left\{H_1(\nu_t)+\left[\frac{\mathcal{M}_{3}}{6}H_4(\nu_t)-\frac{\mathcal{M}_{R}^{(I)}}{2}H_2(\nu_t)-\frac{\mathcal{M}_{R}^{(II)}}{2}\right]\right.\\\nonumber
&&\left.+\frac{\mathcal{M}_4}{24}H_5(\nu_t)-\frac{\mathcal{M}_4^{(I)}}{6}H_3(\nu_t)-\frac{\mathcal{M}_4^{(II)}}{2}H_1(\nu_t)+\dots\right\}\;.
\ea

One might also in principal use Minkowski functionals to analyze fields other than the CMB \cite{Hikage:2006fe, Munshi:2011wu}, and it is straightforward to extend our results here to those scenarios.

\section{Conclusions}
\label{sec:conclusions}

In this paper we have clarified how different classes of inflaton interactions are distinguished by the structure of correlation functions.  We have found that couplings to other (non-inflaton) sectors lead to a very different structure than self-interactions.  This difference may be relevant for LSS probes or, indeed, any observable that is sensitive to information about the fluctuations beyond the 3-point correlation function.  Our analysis also sheds considerable light on the theoretical underpinnings of inflation.  We have clarified the relation (or lack thereof) between the validity of perturbation theory and the Gaussianity of the PDF and uncovered novel new kinds of consistency relations among moments.  Our results indicate that there may be interest in exploring a much wider class of non-Gaussian PDFs than has previously been studied.  Moreover, we have computed the parametric scaling of $n$-point functions generated from inverse decay and also production of massive particles during inflation.

At the technical level, we have shown how to construct the PDF for an arbitrary structure of cumulants and how to translate this into a prediction for the halo mass function and Minkowski functionals appropriate for the feeder scenario.  This construction highlights an important point, which appears not to have been appreciated in previous literature.  Namely, \emph{any} truncated expansion of the PDF about a Gaussian ansatz makes assumptions about the scaling of the moments.  Here we have made these assumptions explicit, showing how different truncations are appropriate for different microphysical inflationary scenarios.

A detection of non-Gaussianity in the equilateral shape template would reveal interesting physics from the inflationary era. However, it 
could not be conclusively ascribed to a single effect (eg, small sound speed). We have demonstrated in this paper that a second possible 
source of nearly equilateral non-Gaussianity (which arises in a very natural way) differs qualitatively from small sound speed models, through the scaling of higher moments.\footnote{Note that the detailed shape of the bispectrum and the value of the tensor-to-scalar ratio in the two cases may also be different \cite{Barnaby:2011vw}.}  Differentiating between a small sound speed and feeder field scenarios would require at least one of three measurements: distinguishing the shapes at the level of the bispectrum, a measurement of the corresponding trispectrum amplitude, or a measurement of the tail of the halo mass function. All three are extremely challenging, but it would be interesting to work out which is most feasible. Finally, we point out that even if differentiating these scenarios is difficult, they demonstrate that the level of non-Gaussianity in the mass function or Minkowski functionals may not correlate with the amplitude of the three-point function in the way that has so far been assumed. Our scenarios show a wider range of possibilities.  This analysis illustrates that great care is required to interpret what the statistics of rare objects imply for the physics of the primordial fluctuations. 

Our construction of a non-hierarchical PDF may have applications beyond the models considered in this paper.  For example, one might consider a phenomenological Ansatz for the primordial curvature perturbation of the form
\begin{equation}
\label{silly}
  \zeta = \zeta_g + \sigma_g^2 - \langle \sigma_g^2\rangle
\end{equation}
where $\zeta_g$ and $\sigma_g$ are statistically independent Gaussian fields.  Such a phenomenological approach has been considered in \cite{Boubekeur:2005fj,Suyama:2008nt}.  (See also \cite{Smith:2011if,Yokoyama:2011qr} for related discussion.)  The Ansatz (\ref{silly}) is expected to give a structure of cumulants similar to the feeder mechanism.  However, being local in position space, Eqn.~(\ref{silly}) may be more easily ammenable to numerical simulations. 

In the body of the paper, we have emphasized that the feeder mechanism is a ``single field" scenario because a single field sources the inflationary background and super-horizon fluctuations, and because there is no possibility of isocurvature. Interestingly, however, there {\it are} two statistically independent contributions to the fluctuations in the power spectrum, as shown in Eq.(\ref{2-pt-scale}). This means that there may be observable stochasticity, since the non-Gaussianity is only correlated with a part of the two-point function (or effectively uncorrelated if the feeder contribution to the two-point is very small). So far, stochasticity has been discussed when the post-inflationary gravitational perturbation has sources from two different primordial fields \cite{Tseliakhovich:2010kf, Smith:2010gx}. In the feeder field case, it is the original inflationary curvature perturbations that exhibit stochasticity, and this will carry over into LSS observables. Finally, we note that quasi-single field scenarios \cite{Chen:2009zp} may also exhibit feeder scaling in some regimes.

The examples which we have studied assume that one starts with the full inflaton action, designed to generate both the nearly de Sitter 
background evolution and the spectrum of fluctuations. A different approach is to separate the physics of the background inflationary 
solution from the physics of the fluctuations - to plead ignorance about anything about the UV mechanism that may relate those points. This 
is the effective field theory of the fluctuations \cite{Cheung:2007st}, which parametrizes single-field non-Gaussianity.  The usual approach 
to the effective field theory of fluctuations in single field inflation \cite{Cheung:2007st} (and see \cite{Bartolo:2010di} for a detailed 
discussion of the trispectrum) gives a hierarchical scaling of moments, by construction.  There is, of course, no conflict with our results 
since the kinds of interactions that give rise to the feeder mechanism are not usually considered (although they might find a natural home 
in the framework established in \cite{Senatore:2010wk}).  One might imagine trying to incorporate analogous effects into the ``single field'' 
framework by replacing the feeder field with an external source.  Such a term would give rise to a tadpole contribution that modifies the Friedman equation and is usually omitted. In the feeder scenarios that we have considered, one can work in a consistent regime where modifications to the Friedman equation are negligible, although there is still a non-trivial impact on cosmological fluctuations. 

In this work we have considered only two realizations of the feeder mechanism, arguing that both give rise to a similar structure of 
correlation functions.  We expect that this structure is quite general.  It would be interesting to construct more explicit models of 
this type and verify this expectation.

\acknowledgments

It is a pleasure to thank Louis Leblond, Marilena LoVerde, Enrico Pajer, Marco Peloso and Lorenzo Sorbo for helpful comments and discussions. We also thank the organizers of the Aspects of Inflation worksop in College Station for providing the forum where this work began. N.B. is supported in part by DOE grant DE-FG0294ER-40823 at UMN.  S.S. is supported by the Perimeter Institute and Eberly Research Funds of The Pennsylvania State University. Research at the Perimeter Institute is supported in part by the Government of Canada through NSERC and by the Province of Ontario through the Ministry of Research and Information (MRI).

\renewcommand{\theequation}{A-\arabic{equation}}
\setcounter{equation}{0}

\appendix
%\section{APPENDIX A: Taxonomy of Inflation Models}
\section{Taxonomy of Inflation Models}
\label{sec:taxonomy}

\subsection{A First Taste of Inflation: Vanilla Flavoured Models}

In the vanilla scenario, the inflaton is effectively decoupled from all other species of particles.  Self interactions of the scalar curvature mode arise either through the inflationary potential or through gravitational couplings.  In both cases, the interaction strength is controlled by slow roll parameters, leading to an important consistency relation between the size of non-Gaussianity and the background dynamics.  That is, the cosmological fluctuations in the vanilla scenario deviate from Gaussianity to roughly the same extent that the background deviates from pure de Sitter \cite{Acquaviva:2002ud,Maldacena:2002vr,Seery:2005wm}.

The vanilla model (\ref{vanilla}) is simple, but not entirely satisfying for several reasons.  The first reason is phenomenological.  The observational signatures of the vanilla model are minimal: we have already observed the the amplitude of the curvature fluctuations (the ratio of the scales $H$ and $\sqrt{\epsilon} M_p$) and have perhaps obtained some evidence for a red spectral index.  The only observable left is the amplitude of gravitational waves (which determines $H$); however, this may well be too small to measure in the next decade.  Primordial non-Gaussianity, although present, will be swamped by post-inflationary gravitational evolution \cite{Pyne:1995bs,Bartolo:2006cu,Bartolo:2006fj,Nitta:2009jp,Pitrou:2008hy} and so is also extremely difficult to measure in the immediate future.  Although the vanilla model is consistent with the data \cite{Komatsu:2010fb}, it will be very hard to say how it can be genuinely tested in the scalar sector.  On the other hand, extensions of the vanilla scenario that give observably large correlations beyond the two-point are much more falsifiable.

The second reason for exploring alternatives to the vanilla scenario is theoretical.  If the basic framework of the inflationary paradigm is correct, it may perhaps be surprising to find only one degree of freedom, decoupled from the rest of particle physics, and no relevant energy scales in the problem besides those associated with gravity or the slow roll potential.  Indeed, efforts to embed inflation within a realistic particle physics framework (such as string theory, the MSSM or supergravity) typically contain many dynamical scalars, in addition to various couplings between the inflationary sector and matter fields.  (We use the word ``matter'' to denote any field which is not involved in driving the inflationary expansion.)  In section \ref{sec:shift} we provided a simple and explicit example illustrating how a rich, non-vanilla interaction structure arises automatically once we try to realize single-field inflation in a more complete particle physics framework.  These ``new'' interactions are controlled by a scale, $f\ll M_p$, and they can lead to observationally interesting non-Gaussianity.

Once we move beyond the vanilla scenario, there are clearly several possibilities: the background evolution and/or the curvature fluctuations on superhorizon scales may have contributions from multiple fields, and the fields sourcing the background may or may not be the same as those sourcing the curvature fluctuations.

\subsection{Classifying Models: I am 32 flavours and then some}

We are now in a position to provide a taxonomy of known inflationary models.  To avoid confusion, we remind the reader that our definition of ``single field'' still allows a ``single field'' inflaton to couple directly to matter fields.

\begin{enumerate}
  
  \item {\bf Decoupled Single Field Inflation}:  There is only a single scalar degree of freedom which sources both the background evolution and also the curvature fluctuations.  The inflaton is effectively decoupled from matter fields during inflation.  Observable non-Gaussianity may be generated if there are self-interactions -- beyond the usual slow-roll/gravitationally suppressed ``vanilla'' couplings -- controlled by a scale $f \ll M_p$.  In this case, the bispectrum is closer to equilateral than local, since it is generated by physical processes on scales of order the horizon or smaller.  There are no iso-curvature perturbations and no super-horizon evolution of $\zeta$.
  
  \item {\bf Coupled Single Field Inflation}: There is a single scalar inflaton degree of freedom and the background dynamics are of the slow roll type.  Nontrivial couplings to matter fields may generate significant non-Gaussianity via the feeder mechanism; see subsection \ref{subsec:feeder}.  The bispectrum is closer to the equilateral than local shape, since the relevant interactions are most significant near horizon crossing.  There are no appreciable large-scale iso-curvature perturbations or super-horizon evolution.
  
  \item {\bf Dissipative Models:} There is a single scalar inflaton degree of freedom which is very strongly coupled to matter fields.  The production of matter quanta is so strong that dissipative effects contribute the dominant source of friction for the motion of the inflaton.  Examples include trapped inflation \cite{Kofman:2004yc,Green:2009ds} (see also \cite{Battefeld:2010sw,Battefeld:2011yj} for generalizations) and the Anber \& Sorbo model \cite{Anber:2009ua}, both of which may be thought of as an extreme limit of the feeder mechanism.  Non-Gaussianity is expected to be closer to equilateral than local.  (Warm inflation \cite{Berera:1995ie} might also be included in this category.)

  \item {\bf Curvaton}: Here the link between the background evolution and the curvature fluctuations is loosened. The fluctuations of a second field, irrelevant during inflation, generate the dominant contribution to the curvature perturbations at the end of inflation. Non-Gaussianity is of the local type and isocurvature modes may be relevant depending on details of the scenario. When the curvaton does not completely dominate the fluctuation spectrum (a mixed inflaton/curvaton model) the perturbation spectrum can be fundamentally multi-field.

  \item {\bf Multi-field inflation}: Here more than one field contributes both to the background evolution and to the fluctuations. Non-Gaussianity is generated by super-horizon evolution and so is of the local type.  Isocurvature modes may be relevant depending on details of the scenario.

\end{enumerate}

See also Table \ref{table:models}.  The categories which we have enumerated here are not meant to exhaust all possibilities.  Rather, these fundamental mechanisms should be understood as basic building blocks which can be assembled in various combinations to construct ever more elaborate models.

\begin{table}[htbp]
{\caption{Classifying inflationary scenarios}\label{table:models}}
  \begin{center}
  \begin{tabular}{|c|c|c|c|c|c|} 
  \hline
  \hline
   Model & Background & Super-horizon $\zeta$ & Non-Gauss. & Iso-Curv. & Single field? \\ 
  \hline
  vanilla & single field & same field & not local & none & $\surd$\\
%    vanilla inflation & single field & same field & not local & none & $\surd$\\
  feeder & single field & same field & not local & negligible& $\surd$ \\
%    feeder mechanism & single field & same field & not local & negligible& $\surd$ \\
  dissipative & single field & same field & not local (?) & negligible (?) & $\surd$ \\
%dissipative models & single field & same field & not local (?) & negligible (?) & $\surd$ \\
  curvaton & single field & different field & local & yes & $\times$ \\
  multi-field & many fields & many fields & local & yes & $\times$ \\
  \hline
  \end{tabular}
  \end{center}
\end{table}

\renewcommand{\theequation}{B-\arabic{equation}}
\setcounter{equation}{0}

%\section{APPENDIX B: Non-Gaussian Correlators from Inverse Decay}
\section{Non-Gaussian Correlators from Inverse Decay}

In subsection \ref{subsec:EFT} we introduced a general effective field theory description (\ref{Ltot}) which we expect to be valid in any model of 
single-field inflation that is characterized by an underlying shift symmetry.  In subsection \ref{subsec:shiftNG} we discussed three distinct mechanisms which operate in this theory, each of which may give rise to observationally interesting non-Gaussianity. The first two mechanisms -- sound speed and resonance effects -- have already received considerable attention in the literature.  The third mechanism, on the other hand, is newer \cite{Barnaby:2010vf}.  In this appendix we compute the parametric scaling of \emph{all} non-Gaussian $n$-point functions generated by this mechanism.  The calculation is a straightforward application of the formalism detailed in \cite{Barnaby:2011vw} and the reader is referred to that paper for more details.

\subsection{Cosmological Perturbation Theory}

Following \cite{Barnaby:2010vf}, we focus on the theory (\ref{Ltot}) in the case where both higher derivative corrections and also resonance effects 
are negligible.  We adopt the flat slicing and Coulomb gauge.  By integrating out the shift and lapse functions one may derive an action for 
the perturbations $\delta\varphi(t,{\bf x}) = \varphi(t,{\bf x}) - \phi(t)$ and $A_\mu(t,{\bf x})$.  Keeping terms up to third order, this 
may be written as \cite{Barnaby:2011vw}:
\begin{equation}
\label{Stot}
  S = S_0 + \delta S_{\mathrm{grav}}
\end{equation}
The leading order action contains the kinetic terms for the gauge field and inflaton, in addition to the pseudo-scalar coupling:
\begin{eqnarray}
  S_0 &=& \int d\tau d^3 x 
                        \left[ \frac{a^2}{2}(\delta\varphi')^2 - \frac{a^2}{2}(\grad \delta\varphi)^2 - \frac{a^2 M^2}{2} (\delta\varphi)^2  
                                - \frac{\alpha}{f}\delta\varphi \epsilon_{ijk} A_i'\partial_jA_k \right] \nonumber \\ \nonumber \\
  &+& \int d\tau d^3x \left[ \frac{1}{2} A_i' A_i' - \frac{1}{2}\partial_iA_j \partial_iA_j \right] \label{S0}
\end{eqnarray}
where we have defined $M^2 \equiv V'' - 3 \left(\frac{\phi'}{a M_p}\right)^2$ and $\partial^{-2}$ is the inverse Laplacian 
satisfying $\partial^{-2}\partial_i \partial^i f = f$.

The full third-order action (\ref{Stot}) also contains cubic interactions which are associated with the nonlinearity of gravity and hence are 
suppressed, both by the Planck scale and also by slow roll parameters.  In (\ref{Stot}) we have denoted these by $\delta S_{\mathrm{grav}}$.  
Explicitly, these are \cite{Malik:2007nd,Seery:2008ms}:
\begin{eqnarray}
 \delta S_{\mathrm{grav}} &=& \int d\tau d^3 x \, \frac{\sqrt{2\epsilon}}{M_p} \, \delta\varphi \, \left( -\frac{1}{4} A_i'A_i' - \frac{1}{8} F_{ij}F_{ij} + \frac{1}{2} \partial^{-2}\partial_\tau\partial_i(F_{ij}A_j')  \right)   \nonumber \\
&+& \int d\tau d^3x \, \frac{\sqrt{2\epsilon}}{M_p} \, \left( \frac{a^2}{2}\delta\varphi'\partial_i\partial^{-2}(\delta\varphi')\partial_i\delta\varphi
                                                       - \frac{a^2}{4}\delta\varphi(\delta\varphi')^2 - \frac{a^2}{4}\delta\varphi(\grad\delta\varphi)^2   \right) \label{Sgrav}
\end{eqnarray}
Evidently, the strength of the interactions encoded in (\ref{Sgrav}) is set by the coupling constant $\frac{\sqrt{\epsilon}}{M_p}$.  This 
should be compared to the scale $\frac{\alpha}{f}$ which controls the strength of the pseudo-scalar interaction in (\ref{S0}).  As discussed 
in section \ref{sec:shift}, the natural expectation from effective field theory is $\alpha=\mathcal{O}(1)$ and $f\ll M_p$.  In this case,
we have $\frac{\alpha}{f} \gg \frac{\sqrt{\epsilon}}{M_p}$ and the gravitational couplings $\delta S_{\mathrm{grav}}$ may be neglected.  
Quantitatively, we expect $\epsilon^{1/2} \sim 10^{-1}$ for a large-field inflation model, whereas $\alpha M_p / f \sim 10^{2}$ whenever
non-Gaussianity is observationally interesting \cite{Barnaby:2010vf}.  Hence, we expect that the gravitational couplings are negligible in the most 
interesting scenario.

Notice that the terms on the second line of (\ref{Sgrav}) are just the usual trilinear scalar interactions which would arising in ordinary
slow roll inflation, even in the absence of the gauge field.  These are known to contribute 
$\Delta f_{NL} = \mathcal{O}(\epsilon,\eta)$ \cite{Acquaviva:2002ud,Maldacena:2002vr,Seery:2005wm,Malik:2007nd,Seery:2008qj}.
It follows that such interactions may be neglected whenever $f_{NL} \gsim \mathcal{O}(1)$.

\subsection{Production of Gauge Fluctuations}

The key aspect of the dynamics of the model (\ref{Stot}) is the production of fluctuations, $\delta A_\mu$, due to the homogeneous motion of 
the inflaton.  To capture this effect we study the equation of motion for the gauge field fluctuations in the homogeneous background 
$\phi(t)$:
\begin{equation}  
  \vec{A}'' - \nabla^2 \vec{A} - \frac{\alpha \phi'}{f} \nabla\times \vec{A} = 0
\end{equation}
The last term arises due to the coupling of the gauge fluctuations to the time-dependent condensate.  This term breaks the conformal invariance
of the gauge field sector and, roughly speaking, becomes important for wave-numbers $k \lsim \alpha \phi' / f$.

We decompose the q-number field $\vec{A}(\tau,{\bf x})$ as 
\begin{equation}
  \vec{A}(\tau,{\bf x}) = \sum_{\lambda=\pm} \int \frac{d^3k}{(2\pi)^{3}} \left[ \vec{\epsilon}_\lambda({\bf k}) a_{\lambda}({\bf k}) A_\lambda(\tau,{\bf k}) e^{i {\bf k}\cdot {\bf x}} + \mathrm{h.c.}   \right]
\label{decomposition}
\end{equation}
where ``$\mathrm{h.c.}$'' denotes the Hermitian conjugate of the preceding term, the annihilation/creation operators obey
\begin{equation}
  \left[a_{\lambda}({\bf k}), a_{\lambda'}^{\dagger}({\bf k'})\right] =(2\pi)^3 \delta_{\lambda\lambda'}\delta^{(3)}({\bf k}-{\bf k'})
\label{ladder}
\end{equation}
and $\vec{\epsilon}_\lambda$ are circular polarization vectors (see \cite{Barnaby:2011vw} for more details).  Using the decomposition 
(\ref{decomposition}) the equation of motion for the c-number mode functions in the classical background $\phi(t)$ takes the form
\begin{equation}
\label{Amode}
  \left[ \frac{\partial^2}{\partial\tau^2} + k^2 \pm \frac{2 k \xi}{\tau} \right] A_{\pm}(\tau,k) = 0, \hspace{5mm} \xi \equiv \frac{\alpha \dot{\phi}}{2 f H}
\end{equation}
During inflation the parameter $\xi$ may be treated as constant, as its time variation is subleading in a slow roll expansion.

From equation (\ref{Amode}) we see that one of the polarization states experiences a tachyonic instability for 
$k/(aH) \lsim 2\xi$. (Notice that this effect could arise also in flat space, the dependence of $\xi$ on the expansion 
rate $H$ arises due through time dependence of $\phi(t)$.)  Without loss of generality, we assume that $\dot{\phi} > 0$ 
during inflation, so that the mode exhibiting the instability is $A_+$.  The correctly normalized solutions of (\ref{Amode}) are well approximated by
\begin{equation}
\label{mode_soln}
  A_{+}(\tau,k) \cong \frac{1}{\sqrt{2k}} \left(\frac{k}{2\xi a H}\right)^{1/4} e^{\pi \xi - 2\sqrt{2 \xi k / (aH)}}
\end{equation}
in the interval $(8\xi)^{-1} \lsim k/(aH) \lsim 2\xi$ of phase space that accounts for most of the power in the produced gauge 
fluctuations.  The phase space of growing modes is non-vanishing for $\xi \gsim \mathcal{O}(1)$, which we assume throughout.  Notice the 
exponential enhancement $e^{\pi \xi}$ in the solution (\ref{mode_soln}).  This arises due to the tachyonic instability and reflects significant 
nonperturbative gauge particle production in the regime $\xi \gsim 1$.  On the other hand, the production of gauge field fluctuations is 
uninterestingly small for $\xi < 1$.  The other polarization state, $A_{-}(\tau,k)$, is not produced and can therefore be 
ignored.

\subsection{Inverse Decay Effects}

In the last subsection, we showed that the homogeneous motion of the inflaton leads to a tachyonic production of gauge fluctuations.  Throughout this paper, we work in the conventional slow roll regime where the backreaction of these produced fluctuations on the homogeneous background is negligible; more on this later.\footnote{See \cite{Anber:2009ua} for an alternative scenario where backreaction effects are important.} Nevertheless, the production of gauge quanta still has an important effect on the observable cosmological fluctuations in the model: the produced gauge fluctuations may source inflaton perturbations via inverse decay, $\delta A + \delta A \rightarrow \delta\varphi$.

We decompose the q-number inflaton fluctuation as 
\begin{equation}
\label{fourier}
  \delta\varphi(\tau,{\bf x}) = \int \frac{d^3k}{(2\pi)^{3}} \frac{Q_{\bf k}(\tau)}{a(\tau)} e^{i {\bf k}\cdot {\bf x}}
\end{equation}
The equation of motion for the Fourier modes is
\begin{eqnarray}
  && \left[\partial_\tau^2 + k^2 + a^2 M^2 - \frac{a''}{a} \right] Q_{\bf k}(\tau) = J_{\bf k}(\tau) \label{phi_cor} \\
  && J_{\bf k}(\tau) \equiv a^3 \frac{\alpha}{4f}  \int d^3x e^{-i{\bf k}\cdot {\bf x}} \left[ F^{\mu\nu} \tilde{F}_{\mu\nu} 
  - \langle F^{\mu\nu}\tilde{F}_{\mu\nu}\rangle \right]
   \label{J}
\end{eqnarray}
where the quantity $F\tilde{F}$ is constructed from the solutions (\ref{mode_soln}).  The solution of (\ref{phi_cor}) may be split into two 
parts
\begin{equation}
  Q_{\bf k}(\tau) = \underbrace{Q_{\bf k}^{\mathrm{vac}}(\tau)}_{\mathrm{homogeneous}} 
  + \underbrace{Q_{\bf k}^{\mathrm{inv.dec}}(\tau)}_{\mathrm{particular}}
\label{hom-inhom2}
\end{equation}
The homogeneous solution corresponds, physically, to the usual vacuum fluctuations from inflation.  The particular solution, on the other hand,
may be interpreted as arising due to inverse decay processes.

The usual vacuum fluctuations from inflation are Gaussian, to a very good approximation \cite{Acquaviva:2002ud,Maldacena:2002vr,Seery:2005wm}.  Hence, the 
homogeneous term may be expanded as
\begin{equation}
\label{hom}
  Q_{\bf k}^{\mathrm{vac}}(\tau) = b({\bf k}) u_k(\tau) + b^\dagger({-{\bf k}}) u_k^\star(\tau)
\end{equation}
where the c-number modes $u_k$ are the well-known homogeneous solutions of (\ref{phi_cor}).  The inflaton ladder operators obey
\begin{equation}
  \left[b({\bf k}), b^\dagger({\bf k'})\right] = (2\pi)^3 \delta^{(3)}({\bf k}-{\bf k'})
\label{ladder-phi}
\end{equation}
and commute with the ladder operators of the gauge field
\begin{equation}
   \left[b({\bf k}), a_\lambda({\bf k'})\right] = \left[b({\bf k}), a^\dagger_\lambda({\bf k'})\right] = 0
\end{equation}

Next, we turn our attention to the particular solution of (\ref{phi_cor}), which may be written as
\begin{equation}
\label{par}
  Q^{\mathrm{inv.dec}}_{\bf k}(\tau) = \int_{-\infty}^{0}d\tau' G_k(\tau,\tau') J_{\bf k}(\tau')
\end{equation}
where the source term was defined in (\ref{J}) and $G_k(\tau,\tau')$ is the retarded Green function which solves
\begin{equation}
  \left[\partial_\tau^2 + k^2 + a^2 \mathcal{M}^2 - \frac{a''}{a} \right] G_{k}(\tau,\tau') = \delta(\tau-\tau')
\end{equation}
The $n$-point correlation functions take the form
\begin{eqnarray}
  && \langle Q^{\mathrm{inv.dec}}_{\bf k_1}(\tau) \cdots Q^{\mathrm{inv.dec}}_{\bf k_n}(\tau) \rangle = \nonumber \\ 
  && \left[ \prod_{i=1}^n \int d\tau_i G_{k_i}(\tau,\tau_i)\right] \times  \langle J_{k_1}(\tau_1) \cdots J_{k_n}(\tau_n) \rangle
\end{eqnarray}

Ultimately, we are interesting in the co-moving curvature perturbation, $\zeta(\tau,{\bf x})$, which is related to the field fluctuation as (\ref{convert}).  Working in Fourier space, the contribution to $\zeta_{\bf k}$ from inverse decay effects is
\begin{equation}
  \zeta_{\bf k}^{\mathrm{inv.dec}}(\tau) = -\frac{H}{a \dot{\phi}} Q_{\bf k}^{\mathrm{inv.dec}}(\tau)
\end{equation}
Following closely the formalism detailed in \cite{Barnaby:2011vw}, it can be shown that, on large scales $-k\tau \ll 1$, this may be written
in a suggestive form:
\begin{eqnarray}
  \label{zeta_inv.dec}
  \zeta_{\bf k}^{\mathrm{inv.dec}}(\tau) &\cong& 0.0031\, \frac{H^2}{2\pi|\dot{\phi}|} \frac{\alpha H}{f} \left(\frac{e^{2\pi\xi}}{\xi^4}\right) \frac{1}{k^3} \,
  \int \frac{d^3 q}{(2\pi)^3} \,\, \psi(\hat{k}, \vec{q}_\star)
  \nonumber \\
  &&
     \times \left[ a_+({\bf q})a_+({\bf k-q}) + a^\dagger_+({\bf -q}) a_+({\bf k-q}) + a^\dagger_+({\bf -k+q})a_+({\bf q}) 
           \right.  \\\nonumber
             &&\left.\,\,\,\,\,\,\,\,\, + a^\dagger_+({\bf -q}) a^\dagger_+({\bf -k + q})   \right]
\end{eqnarray}
Here we have introduced dimensionless ratios $\hat{k} \equiv \vec{k}/k$, $\vec{q}_\star \equiv \vec{q}/k$ and also defined the quantity
\begin{equation}
  \psi(\hat{k}, \vec{q}_\star) \cong 4.2\cdot 10^{2} \, 
  \epsilon_i(\vec{q}_\star)\epsilon_i(\hat{k} - \vec{q}_\star) \, \frac{ |\vec{q}_\star|^{1/4} |\hat{k} - \vec{q}_\star|^{1/4} }{ \left( |\vec{q}_\star|^{1/2}+|\hat{k} - \vec{q}_\star|^{1/2}  \right)^{7} }
\end{equation}
which is dimensionless and has no dependence of model parameters whatsoever.  We have normalized the $\psi$-function so that
\begin{equation}
  \int \frac{d^3q_\star}{(2\pi)^3} |\psi(\hat{k},\vec{q}_\star)|^2 = \frac{1}{2}
\end{equation}
which cancels the combinatorial factor arising from the Wick contractions in the 2-point correlator.

Equation (\ref{zeta_inv.dec}) has been written in a way which makes manifest the dependence of $\zeta_{\bf k}^{\mathrm{inv.dec}}(\tau)$ on 
model parameters.  The prefactor multiplying (\ref{zeta_inv.dec}) can be understood as follows.  The factor $\frac{H^2}{2\pi|\dot{\phi}|}$,
which would give the amplitude of fluctuations in the vanilla scenario, can be traced to the relation (\ref{convert}).  The quantity 
$\frac{\alpha H}{f}$, on the other hand, is the natural dimensionless measure of the strength of the pseudo-scalar interaction that gives rise
to inverse decay effects.  Finally, the contribution $\frac{e^{2\pi\xi}}{\xi^4}$ is associated with the time dependence of the gauge modes 
(\ref{mode_soln}).\footnote{In order to derive (\ref{zeta_inv.dec}) we work in the limit $\xi \gg 1$.  At smaller values of $\xi$ the power-law dependence of this term will be modified, as will the factors order unity.  Neither of these modifications will have any impact on our qualitative results for the structure of the correlation functions.}  This last term depends on $\dot{\phi}$, $H$ and $\alpha/f$, but only through the dimensionless combination $\xi$ that controls the effective mass of the gauge field; see equation (\ref{Amode}).

Two things are evident from equation (\ref{zeta_inv.dec}).  First, the particular solution (\ref{par}) is statistically independent of the 
homogeneous solution (\ref{hom}).  Indeed, the former can be expanded in terms of the annihilation/creation operators 
$a_\lambda({\bf k}),a_\lambda^\dagger({\bf k})$ associated with the gauge field while the latter is expanded in terms of the 
annihilation/creation operators $b({\bf k}),b^\dagger({\bf k})$ associated with the inflaton vacuum fluctuations. As we  pointed out,
 these two sets of operators commute with one another.

The second notable feature of equation (\ref{zeta_inv.dec}) is its non-Gaussianity, which is evident from the quadratic dependence on the 
annihilation/creation operators $a_\lambda({\bf k}),a_\lambda^\dagger({\bf k})$.  This is easily understood: the particular solution 
$Q_{\bf k}^{\mathrm{inv.dec}}$ is bi-linear in the Gaussian field $\delta A_\mu$ and hence we expect this to behave, heuristically, as the 
square of a Gaussian field.

\subsection{Parametric Scaling of the Correlation Functions}

We now turn our attention to the $n$-point functions of the co-moving curvature perturbation
\begin{equation}
  \zeta_{\bf k}(\tau) 
  = -\frac{H}{a\dot{\phi}} \left[ Q_{\bf k}^{\mathrm{vac}}(\tau) + Q_{\bf k}^{\mathrm{inv.dec}}(\tau)  \right]
\end{equation}
Before considering non-Gaussian correlators, the 2-point function deserves special attention.  We have \cite{Barnaby:2010vf,Barnaby:2011vw}
\begin{eqnarray}
  \langle \zeta_{\bf k} \,  \zeta_{\bf k'} \rangle &=& 
                                           \frac{H^2}{a^2 \dot{\phi}^2} \left[ 
                                           \langle Q^{\mathrm{vac}}_{\bf k} \, Q^{\mathrm{vac}}_{\bf k'} \rangle   
                                         + \langle Q^{\mathrm{inv.dec}}_{\bf k} \, Q^{\mathrm{inv.dec}}_{\bf k'} \rangle \right] \nonumber \\
 &=& \frac{H^4}{(2\pi)^2|\dot{\phi}|^2} 
     \left[ 2\pi^2 + \left(0.0031\cdot10^{-4} \frac{\alpha H}{f} \frac{e^{2\pi\xi}}{\xi^4} \right)^2  \right] 
     \frac{(2\pi)^3}{k^3} \delta^{(3)}\left({\bf k}+{\bf k'}\right)
\end{eqnarray}
Notice that there is no ``cross-term'' because the two contributions to $\zeta_{\bf k}$ are statistically independent.  
The first term in the square braces comes from the quantum vacuum fluctuations, while the second term arises due to inverse decay effects.

We now turn our attention to a parametric computation of all connected $n$-point functions, with $n \geq 3$.  These correlators encode 
departures from Gaussianity, whereas the usual vacuum fluctuations from slow roll inflation are known to be Gaussian to high accuracy
\cite{Acquaviva:2002ud,Maldacena:2002vr,Seery:2005wm}.  Hence, it will be a good approximation to consider only the particular solution of (\ref{phi_cor})
and neglect the homogeneous solution.  In this approximation, and in the limit $-k\tau \ll 1$, we can write
\begin{eqnarray}
  \langle \zeta_{\bf k_1} \zeta_{\bf k_2} \cdots \zeta_{\bf k_n} \rangle_c &=& 
  \left(0.0031 \frac{H^2}{2\pi|\dot{\phi}|} \frac{\alpha H}{f}  \frac{e^{2\pi\xi}}{\xi^4}\right)^n \, \frac{1}{k^{3n}} \nonumber \\
  && \times \left[ \int \prod_{i=1}^n \frac{d^3 q_i}{(2\pi)^3} \,\, \psi(\hat{k}_i,\vec{q}_{\star,i}) \right] \times \langle  a_{\bf q_1} a_{\bf k_1-q_1} \left[ \cdots \right] a^\dagger_{\bf -q_n} a^\dagger_{\bf -k_n+q_n}  \rangle
\label{scaling_intermediate}
\end{eqnarray}
where we have employed equation (\ref{zeta_inv.dec}).  The expectation value on the last line involves a sum of products of $2n$ 
annihilation/creation operators.  On evaluating the Wick contractions, these give $n$ delta functions, each with a factor of $(2\pi)^3$.  By translational invariance, one of these must be $(2\pi)^3\delta^{(3)}\left(\sum_i {\bf k_i}\right)$.  The remaining $n-1$ delta functions kill off all but one of the integrals $\int \frac{d^3 q_i}{(2\pi)^3}$.  The final integral over internal momenta gives a factor $k^{3}$ when put into dimensionless form: $\int \frac{d^3q}{(2\pi)^3} = k^3 \int \frac{d^3q_\star}{(2\pi)^3}$.  Thus, the final correlator must scale like
\begin{equation}
\label{scaling_final}
  \langle (\zeta_k)^n \rangle_c = \frac{A_n}{k^{3(n-1)}} 
  \left(0.0031 \frac{H^2}{2\pi|\dot{\phi}|} \frac{\alpha H}{f} \frac{e^{2\pi \xi}}{\xi^4}\right)^n (2\pi)^3 \delta^{(3)}\left(\sum_i {\bf k_i}\right)
\end{equation}
The factor $k^{-3(n-1)}$ simply reflects the scale invariance of the produced fluctuations.  The coefficients $A_n$ are dimensionless numbers, independent of model parameters, which encode the integration over $\psi$-functions and also combinatorial factors from the Wick contractions.  Our normalization of the $\psi$-function has been chosen so that $A_2=1$ and it may be checked explicitly that $A_3 \approx 2.2$.  We expect that $A_n = \mathcal{O}(1)$ for higher order $n$ also.  Note that at very large values of $n$, combinatorial factors may become important and influence the scaling.  This issue, which is by no means special to the inverse decay scenario, is left to future studies. 

For later convenience, we write our key results in configuration space:
The variance is given by
\begin{equation}
\label{pwr}
  \langle \zeta^2({\bf x}) \rangle = \int \frac{dk}{k} \frac{H^4}{(2\pi)^2|\dot{\phi}|^2} 
     \left[ 1 + \frac{1}{2\pi^2}\left(0.0031 \frac{\alpha H}{f} \frac{e^{2\pi\xi}}{\xi^4} \right)^2  \right] 
\end{equation}
The cumulants are given by
\begin{eqnarray}
  && \langle \zeta^n({\bf x}) \rangle_c = \int \frac{d^3k_1}{(2\pi)^3}\frac{d^3k_2}{(2\pi)^3}\cdots\frac{d^3k_n}{(2\pi)^3} \nonumber \\ 
 &&  \,\,\,\,\,\,\,\times
 \left[(2\pi)^3A_n \left(0.0031 \frac{H^2}{2\pi|\dot{\phi}|} \frac{\alpha H}{f} \frac{e^{2\pi \xi}}{\xi^4}\right)^n 
 \frac{1}{k^{3(n-1)}} \delta^{(3)}\left(\sum_i {\bf k_i}\right)  \right]  \label{inv.dec_cumulant}
\end{eqnarray}

\subsection{Computational Control}

A number of simplifying assumptions were made in our derivation of the key result (\ref{scaling_final}) for the parametric scaling of the 
non-Gaussian correlation functions from inverse decay effects.  In this subsection, we justify those approximations in more detail and draw
attention to the various constraints on model parameters that are imposed by the requirement of computational control.  

As discussed in the last section, we require $H \ll f \ll M_p$ in order for the effective field theory description (\ref{Stot}) to be valid.
In addition to this, we have made two main simplifying assumptions in deriving (\ref{scaling_final}).  Namely:
\begin{enumerate}
  \item {\bf Neglect of Gravitational Interactions:} We have neglected the gravitational interactions encoded in $\delta S_{\mathrm{grav}}$;
  see equation (\ref{Sgrav}).
  \item {\bf Neglect of Backreaction:} We have neglected the backreaction of the produced gauge fluctuations on the homogeneous
  dynamics of $\phi(t)$ and $a(t)$.
\end{enumerate}
We have already argued that the former assumption is justified whenever $\frac{\alpha}{f} \gg \frac{\epsilon^{1/2}}{M_p}$; see the discussion
below (\ref{Sgrav}).  The latter condition, on the other hand, requires somewhat more care.

There are two distinct backreaction effects.  First, the gauge fluctuations $\delta A_\mu$ are produced at the expense of the kinetic energy
of the inflaton condensate, $\phi(t)$.  This effect is encoded in the mean field equation
\begin{equation}
 \ddot{\phi} + 3 H \dot{\phi} + V'(\phi) = \frac{\alpha}{f} \langle \vec{E}\cdot \vec{B} \rangle 
\end{equation}
where $\vec{E} = -\frac{1}{a^2} \vec{A}'$ and $\vec{B} = \frac{1}{a^2} \vec{\nabla}\times\vec{A}$ are the physical electric and magnetic components of $A_\mu$.  In order to trust the standard slow roll result $3H\dot{\phi} \cong -V'$ we require $|V'| \gg \frac{\alpha}{f} \langle \vec{E}\cdot \vec{B} \rangle$.  This condition may be shown to be equivalent to \cite{Barnaby:2011vw}
\begin{equation}
\label{back}
  0.0031\frac{H^2}{2\pi|\dot{\phi}|} \frac{\alpha H}{f} \frac{e^{2\pi \xi}}{\xi^{4}} \ll \mathcal{O}(1)
\end{equation}

The second backreaction effect arises because the produced gauge fluctuations contribute to the total energy density of the universe, modifying the Friedman equation:
\begin{equation}
  3 H^2 = \frac{1}{M_p^2}\left[ \frac{1}{2}\dot{\phi}^2 + V(\phi) + \frac{1}{2}\langle \vec{E}^2 + \vec{B}^2 \rangle  \right]
\end{equation}
It can be shown that the energy density in gauge field fluctuations is negligible, $\langle \vec{E}^2 + \vec{B}^2 \rangle \ll V$, whenever the condition (\ref{back}) is satisfied.

Notice that the backreaction bound (\ref{back}) constraints the \emph{same} parameter which controls the size of the non-Gaussian $n$-point correlation functions; see equation (\ref{scaling_final}).  Indeed, this condition could be re-phrased as
\begin{equation}
  \langle \zeta_{\mathrm{inv.dec}}^2 \rangle^{1/2} \ll 1
\end{equation}
This is not surprising since, after all, we expect backreaction effects to be negligible when fluctuations are small.

Our solution (\ref{mode_soln}) for the produced gauge fluctuations was obtained neglecting inhomogeneities of $\varphi$.  As a final consistency check, let us verify that this approximation is valid, even in the inverse decay dominated regime.  The equations of motion of the electromagnetic field can be written in the form
\begin{eqnarray}
 && \vec{E}' + 2 a H \vec{E} - \vec{\nabla}\times\vec{B} = -\frac{\alpha}{f} \varphi' \vec{B} - \frac{\alpha}{f} \vec{\nabla}\varphi\times\vec{E} \\
 && \vec{\nabla}\cdot \vec{E} = -\frac{\alpha}{f} \left(\vec{\nabla}\varphi\right) \cdot\vec{B}
\end{eqnarray}
An interative solution of the equations of motion will converge provided
\begin{equation}
 H \, \langle \vec{E}^2 \rangle^{1/2} \gg 
 \frac{\alpha}{f} \, \Big\langle \left(\frac{\vec{\nabla}}{a}\delta\varphi\times\vec{E}\right)\cdot \left(\frac{\vec{\nabla}}{a}\delta\varphi\times\vec{E}\right) \Big\rangle^{1/2}
\end{equation}
We can estimate
\begin{eqnarray}
 \frac{\alpha}{f}\, \Big\langle \left(\frac{\vec{\nabla}}{a}\delta\varphi\times\vec{E}\right)\cdot \left(\frac{\vec{\nabla}}{a}\delta\varphi\times\vec{E}\right) \Big\rangle^{1/2}
 &\sim& \frac{\alpha}{f}\, \Big\langle \left(\frac{\vec{\nabla}}{a}\delta\varphi\right)^2 \Big\rangle^{1/2} \, \langle \vec{E}^2 \rangle^{1/2} \nonumber \\
 &\sim& \frac{\alpha H}{f}\, \langle (\delta\varphi)^2 \rangle^{1/2} \, \langle \vec{E}^2 \rangle^{1/2} \nonumber \\
 &\sim& \xi \, \langle \zeta^2 \rangle^{1/2} \,\,  H \,\langle \vec{E}^2 \rangle^{1/2}
\end{eqnarray}
Hence, we arrive at the condition $\langle \zeta^2\rangle^{1/2}\ll \xi^{-1}$, which is easily satisfied whenever backreaction effects are small (recall that $\xi \gsim \mathcal{O}(1)$ in the most interesting region of paramter space).

\renewcommand{\theequation}{C-\arabic{equation}}
\setcounter{equation}{0}

%\section{APPENDIX C: Non-Gaussian Correlators from Rescattering}
\section{Non-Gaussian Correlators from Rescattering}

In this appendix we estimate the cumulants for the model (\ref{Lresc}).  Decomposing the q-number inflaton fluctuations as (\ref{fourier}) the equation of motion (\ref{resc_eqn}) takes the form
\begin{eqnarray}
  && \left[ \partial_\tau^2 + k^2 + a^2 m^2 -\frac{a''}{a} \right] Q_{\bf k}(\tau) = J_k(\tau) \label{mode_resc} \\
  && J_k(\tau) \equiv - a^3 g k_\star^2 t(\tau) \int d^3x e^{-i{\bf k}\cdot {\bf x}}\left[ \chi^2(\tau,{\bf x}) 
  - \langle\chi^2(\tau,{\bf x}) \rangle  \right]\label{Jresc}  
\end{eqnarray}
where $k_\star \equiv \sqrt{g |\dot{\phi}|}$, $t(\tau) = \int^\tau a(\tau')d\tau'$ and we have arbitrarily set the origin of time so that $\phi(t)=\phi_0$ at $t=0$.  The solution of (\ref{mode_resc}) may be written as
\begin{equation}
  Q_{\bf k}(\tau) = \underbrace{Q_{\bf k}^{\mathrm{vac}}(\tau)}_{\mathrm{homogeneous}} 
  + \underbrace{Q_{\bf k}^{\mathrm{resc}}(\tau)}_{\mathrm{particular}}
\label{hom-inhom3}
\end{equation}
The homogeneous solution corresponds, physically, to the usual vacuum fluctuations from inflation.  This is given explicitly by (\ref{hom}).  The particular solution, on the other hand, may be interpreted as arising due to rescattering.  

The vacuum fluctuations are Gaussian to very high accuracy, hence, to compute the cumulants with $n\geq 3$ we can focus only on the particular solution of (\ref{mode_resc}).  The correlation function of the curvature perturbation is
\begin{equation}
\label{resc_cor_step}
 \langle  \zeta_{\bf k_1}\zeta_{\bf k_2}\cdots \zeta_{\bf k_n}(\tau)\rangle_c = \left(-\frac{H}{\dot{\phi}}\right)^{n}
  \prod_{i=1}^n\int d\tau_i \frac{G_{k_i}(\tau,\tau_i)}{a(\tau)} \times \langle J_{k_1}(\tau_1) J_{k_2}(\tau_2) \cdots J_{k_n}(\tau_n) \rangle 
\end{equation}
Ref.~\cite{Barnaby:2010ke} provides a detailed account of $\chi$-particle production in an expanding universe, showing explicitly how to solve for the produced fluctuations, $\chi(\tau,{\bf x})$.  For our purposes, it is sufficient to note that the $n$-point correlator of the source term can be estimated as
\begin{equation}
\label{Jresc2}
 \langle J_{k_1}(t_1) J_{k_2}(t_2) \cdots J_{k_n}(t_n) \rangle \sim g^n k_\star^3 e^{-\pi c k^2 / (k_\star^2)}
 \, \times \, \prod_{i=1}^n \Theta(t_i) \,\times\, \delta^{(3)}\left(\sum_i{\bf k_i}\right)
\end{equation}
where $c$ is some order unity number.  The product over step functions enforces the fact that the source $J_{k_i}(t_i)$ turns on only for $t_i>0$, after particle production has occured.  The estimate (\ref{Jresc2}) would recieve order unity corrections in a more careful computation, however, this expression correctly captures the parametric dependence and will be quite sufficient for our purposes.

Using the estimate (\ref{Jresc2}) the $n$-point correlation function becomes
\begin{eqnarray}
  \langle  \zeta_{\bf k_1}\zeta_{\bf k_2}\cdots \zeta_{\bf k_n}(\tau) \rangle_c &\sim&
 k_\star^3 \left(g \frac{H^2}{\dot{\phi}}\right)^n \, e^{-\pi c k^2 / (2k_\star^2)} \nonumber \\
 && \,\,\,\, \times \,\left[ \prod_{i=1}^n \int_{\tau_0}^\tau d\tau_i\frac{G_k(\tau,\tau_i)}{a(\tau)} \right]
 \,\, \times \delta^{(3)}\left(\sum_i{\bf k_i}\right)
\end{eqnarray}
where $\tau_0$ is the value of the conformal time variable when $\phi=\phi_0$.  If we set $a = 1$ at $t=0$ then $\tau_0=-1/H$.  The time dependence arises through integrals (called $I_2$ in \cite{Barnaby:2010ke}) of the form
\begin{equation}
 \int_{-1/H}^\tau d\tau_i\frac{G_k(\tau,\tau_i)}{a(\tau)} \approx \frac{1}{H^2} f\left(\frac{k}{H}\right) 
\end{equation}
where we take the limit $-k\tau \rightarrow 0$ and introduce the function
\begin{equation}
  f(x) = \frac{1}{x^3} \int_{0}^{x} \frac{dx'}{x'} \left[\sin(x') - x' \cos(x') \right] 
  = \frac{1}{x^3}\left[ \mathrm{Si}(x) - \sin (x)  \right]
\end{equation}
Here $\mathrm{Si}(x)$ is the sine integral.  

Putting everything together, we estimate the momentum space correlator as
\begin{eqnarray}
 \langle  \zeta_{\bf k_1}\zeta_{\bf k_2}\cdots \zeta_{\bf k_n}(\tau) \rangle_c &\sim&
 \frac{k_\star^3}{H^{3n}} \left( g \,  \frac{H^2}{\dot{\phi}}\right)^n e^{-\pi c k^2 / (2k_\star^2)} \nonumber \\ 
 && \,\,\,\,\,\,\,\,\, \times \, \left[ \prod_{i=1}^n f(k_i/H) \right]
 \, \times \delta^{(3)}\left(\sum_i{\bf k_i}\right)
\end{eqnarray}
As expected, the fluctuations from inverse decay are not scale invariant, rather, the effect peaks near scales leaving the horizon when particle production occured.

The cumulants are
\begin{equation}
 \langle \zeta^n({\bf x}) \rangle = \int \frac{d^3k_1}{(2\pi)^{3}} \,\, \frac{d^3k_2}{(2\pi)^{3}} \cdots \frac{d^3k_n}{(2\pi)^{3}}
  \langle  \zeta_{\bf k_1}\zeta_{\bf k_2}\cdots \zeta_{\bf k_n} \rangle
\end{equation}
Most of the support for this integral comes from scales $k \sim a_0 H$, corresponding to the location of the non-Gaussian feature.  (Recall that we have set $a_0 \equiv a(t=0) = 1$.)  We can smooth the fluctuations on the scale of the feature simply by constraining the domain of integration to lie close to the peak of the integrand.  Proceeding in this way gives
\begin{equation}
\label{resc_c_1}
 \langle \zeta^n({\bf x}) \rangle_{R_b} \sim a_n \frac{k_\star^3}{H^3} \left[ (2\pi g) \, \frac{H^2}{2\pi|\dot{\phi}|}\right]^n
\end{equation}
where the coefficients $a_n$ are independent of model parameters and $R_b \sim H^{-1}$.  The variance recieves important contributions both 
from rescattering effects and also from the usual vacuum fluctuations.  Summing up both contributions, we have
\begin{equation}
\label{resc_c_2}
 \langle \zeta^2({\bf x}) \rangle_{R_b} \sim \frac{H^4}{(2\pi)^2\dot{\phi}^2}\left[ 1 + a_2 (2\pi g)^2 \frac{k_\star^3}{H^3} \right]
\end{equation}
Both equations (\ref{resc_c_1}) and (\ref{resc_c_2}) are very similar to the analogous result for inverse decay effects.  This is not 
surprising, given the qualitative similarity of the underlying physics.

%\bibliography{NGaxion}{}
%\bibliographystyle{JHEP}
%%\bibliographystyle{plain}

\providecommand{\href}[2]{#2}\begingroup\raggedright\endgroup

\end{document}